\documentclass{aastex63}

\usepackage{amsmath}

\newcommand\lya{$\mathrm{Lyman}\,\alpha$}
\newcommand\megam{Mega-MUSCLES}
\newcommand\musc{MUSCLES}
\newcommand{\Msun}{\mbox{$\mathrm{M}_{\odot}$}}

\newcommand{\mearth}{$\rm M_{\earth}$}

\newcommand{\Teff}{\mbox{$T_{\mathrm{eff}}$}}
\newcommand{\logg}{\mbox{$\log g$}}

\graphicspath{{./}{figures/}}

\received{October 8, 2024}
\revised{November 1, 2024}
\accepted{November 5, 2024}
\submitjournal{ApJ}

\shorttitle{Mega-MUSCLES Treasury Survey}
\shortauthors{Wilson et al.}

\begin{document}

\title{The Mega-MUSCLES Treasury Survey: X-ray to infrared Spectral Energy Distributions of a representative sample of M dwarfs}

\correspondingauthor{David J. Wilson}
\email{djwilson394\@gmail.com}

\author[0000-0001-9667-9449]{David J. Wilson}
\affil{Laboratory for Atmospheric and Space Physics, University of Colorado, 600 UCB, Boulder, CO 80309}

\author[0000-0001-8499-2892]{Cynthia S. Froning}
\affiliation{Southwest Research Institute, 6220 Culebra Road, San Antonio, TX 78238, USA}

\author[0000-0002-7119-2543]{Girish M. Duvvuri}
\affil{Department of Physics and Astronomy, Vanderbilt University, Nashville, TN 37235, USA}

\author[0000-0002-1176-3391]{Allison Youngblood}
\affil{NASA Goddard Space Flight Center, Greenbelt, MD 20771}

\author[0000-0002-1002-3674]{Kevin France}
\affil{Department of Astrophysical and Planetary Sciences, University of Colorado, Boulder, CO 80309, USA}
\affil{Laboratory for Atmospheric and Space Physics, University of Colorado, 600 UCB, Boulder, CO 80309}

\author[0000-0003-2631-3905]{Alexander Brown}
\affil{Center for Astrophysics and Space Astronomy, University of Colorado, 389 UCB, Boulder, CO 80309}

\author[0000-0002-5094-2245]{P.\ Christian Schneider}
\affil{Hamburger Sternwarte, Gojenbergsweg 112, 21029 Hamburg }

\author[0000-0002-3321-4924]{Zachory Berta-Thompson}
\affil{Department of Astrophysical and Planetary Sciences, University of Colorado, Boulder, CO 80309, USA}

\author[0000-0001-9546-9044]{Andrea P. Buccino}
\affil{ Dpto. de Física, Facultad de Ciencias Exactas y Naturales (FCEN), Universidad de Buenos Aires (UBA), Buenos Aires, Argentina}





\author[0000-0003-4446-3181]{Jeffrey Linsky}
\affil{JILA, University of Colorado and NIST, Boulder, CO 80309-0440 USA}

\author[0000-0001-5646-6668]{R.~O. Parke Loyd}
\affil{Eureka Scientific, Inc., Oakland, CA 94602, USA}


\author[0000-0002-0747-8862]{Yamila Miguel}
\affil{Leiden Observatory, P.O.\ Box 9500, 2300 RA Leiden, The Netherlands}
\affil{SRON Netherlands Institute for Space Research, Niels Bohrweg 4, 2333 CA Leiden, The Netherlands}

\author[0000-0003-4150-841X]{Elisabeth Newton}
\affil{Department of Physics, Massachusetts Institute of Technology, Cambridge, MA 02139, USA}

\author[0000-0002-4489-0135]{J. Sebastian Pineda}
\affil{Department of Astrophysical and Planetary Sciences, University of Colorado, Boulder, CO 80309, USA}

\author[0000-0003-3786-3486]{Seth Redfield}
\affil{Wesleyan University, Department of Astronomy and Van Vleck Observatory, 96 Foss Hill Dr., Middletown, CT 06459, USA}

\author[0000-0002-2989-3725]{Aki Roberge}
\affil{NASA Goddard Space Flight Center, Greenbelt, MD 20771}

\author[0000-0003-1620-7658]{Sarah Rugheimer}
\affil{Department of Physics and Astronomy, York University, 4700 Keele Street, Toronto, ON M3J 1P3, Canada}


\author[0000-0003-4615-8746]{Mariela C. Vieytes}
\affil{Instituto de Astronomía y Física del Espacio (CONICET-UBA), Buenos Aires, Argentina}


\begin{abstract}
We present 5--$1\times10^7$\,\AA\  spectral energy distributions (SEDs) for twelve M\,dwarf stars covering spectral types M0--M8. Our SEDs are provided for community use as a sequel to the Measurements of the Ultraviolet Spectral Characteristics of Low-mass Exoplanetary Systems (MUSCLES) survey. The twelve stars include eight known exoplanet hosts and four stars chosen to fill out key parameter space in spectral type and rotation period. The SEDs are constructed from Hubble Space Telescope ultraviolet spectroscopy and XMM Newton, Chandra and/or Swift X-ray observations and completed with various model data, including \lya\ reconstructions, PHOENIX optical models, APEC coronal models and Differential Emission Measure models in the currently-unobservable Extreme Ultraviolet. We provide a complete overview of the Mega-MUSCLES program, including a description of the observations, models, and SED construction. The SEDs are available as MAST High-Level Science Products and we describe the various data products here. We also present ensemble measurements from our sample that are of particular relevance to exoplanet science, including the high-energy fluxes in the habitable zone and the FUV/NUV ratio. Combined with MUSCLES, Mega-MUSCLES provides SEDs covering a wide range of M\,dwarf spectral types and ages such that suitable proxies for any M dwarf planet host of interest may be found in our sample. However, we find that ultraviolet and X-ray fluxes can vary even between stars with similar parameters, such that observations of each exoplanet host star will remain the gold standard for interpreting exoplanet atmosphere observations.

\end{abstract}



\section{Introduction} \label{sec:intro}
M\,dwarf stars, with masses $\approx 0.1-0.5$\,\Msun, have emerged as the premier targets for exoplanet discovery and characterization over the past two decades \citep{tarteretal07-1}, particularly for rocky planets with masses and radii comparable to that of Earth in orbits within the habitable zone around the host star. The focus on M\,dwarfs is partly due to their great abundance, representing over 70\,\% of stars \citep{henryetal06-1}, but mostly for observational considerations. The sensitivity of both of the dominant techniques for exoplanet discovery, the transit and radial velocity methods, scale dramatically and favourably with decreased stellar mass, such that planets with similar mass and radius to the Earth are much easier to find around M\,dwarfs than around larger stars. Additionally, the habitable zones of M\,dwarfs are much closer in compared to the habitable zones around larger stars due to their lower luminosity, so the orbital periods of planets there are of order days to weeks rather than years, requiring less time-on-target to detect and confirm. Short orbital periods, and a relatively small planet-to-star size ratio, also makes M\,dwarf planets favourable targets for atmospheric characterisation via transit spectroscopy. The importance of M\,dwarfs to exoplanet science has been highlighted by some of the most remarkable discoveries, such as the habitable zone planet orbiting Proxima Centauri \citep{anglada-escudeetal16-1} and the seven Earth-sized planets around TRAPPIST-1 \citep{gillonetal16-1, gillonetal17-1}. The launch of JWST has enabled an increasing number of atmospheric observations of planets orbiting M\,dwarfs, ruling out thick atmospheres around several of the most accessible targets \citep{greenetal23-1, ziebaetal23-1, lincowskietal23-1, moranetal23-1}. With the advent of the Rocky Worlds program \citep{redfieldetal24-1}, continued JWST observations of planets around M\,dwarfs are guaranteed for years to come.     

When attempting to assess the characteristics and potential habitability of such planets, we must be cautious when comparing them with the Solar system \citep{rugheimeretal15-2, rugheimer+kaltenegger18-1}. The differences between M\,dwarfs and Sun-like stars are many. Their lower temperature means that the spectrum is shifted into the red, and the average M\,dwarf is much more active than more massive stars, both in terms of flare frequency and maximum flare strength relative to its luminosity \citep{loydetal18-1, froningetal19-1}, with the potential effects on their planets exacerbated by the close-in habitable zones \citep{buccinoetal07-1, vidaetal17-1}.

Perhaps the starkest difference between M\,dwarfs and Sun-like stars are their high energy spectral energy distributions (SEDs). Multiple observations have demonstrated that the X-ray fluxes of M\,dwarfs relative to their bolometric luminosities can be thousands of times higher than for Sun-like stars \citep{wheatleyetal17-1, wrightetal18-1, brownetal23-1}. Conversely, the red-ward shift of the photospheric spectrum results in relative near-ultraviolet flux ratios being lower than Sun-like stars by a similar factor. The X-ray and ultraviolet fluxes are key drivers of exoplanet atmospheric structure and chemistry. Photochemistry in the upper atmospheres of exoplanets (i.e., the layers that we can most readily observe) is governed by the incident ultraviolet spectrum, and in particular the strength of the \lya\ line and the ratio of FUV to NUV fluxes \citep{seguraetal05-1, mosesetal13-1, migueletal15-1, jwsters23-1}. X-ray and EUV radiation drives atmospheric escape, such that planets in extreme XUV environments may not retain atmospheres at all \citep{watsonetal81-1, poppenhaegeretal24-1, vanlooverenetal24-1}. 

A number of surveys  have been pursued over the last several years to use HST and other facilities to obtain energetic spectra of M dwarf stars. Programs like Living with a Red Dwarf, HAZMAT, \musc, and FUMES, plus observations focused on individual targets of interest, have greatly expanded the sample of M stars with observed ultraviolet spectra \citep{guinanetal16-1,loydetal21-1,loydetal18-2,franceetal16-1,youngbloodetal16-1,loydetal16-1,loydetal18-1,bourrieretal17-1,macgregoretal21-1,waalkesetal19-1,diamond-loweetal22-1, pinedaetal21-2, feinsteinetal22-1, rockcliffeetal21-1}.

Ideally, a full X-ray through optical SED should be obtained for any M\,dwarf of interest, but such observations are often impractical for multiple reasons. Whilst the high energy emission from M\,dwarfs is a higher fraction of the bolometic flux relative to a larger star, it is still intrinsically faint, requiring large investments of limited space telescope observing time for even the closest M\,dwarfs. Accordingly, the \musc\ Treasury Survey was conceived to provide complete SEDs for a representative sample of low mass stars, while also providing scaling relations for other surveys with more targets but less wavelength coverage \citep{youngbloodetal17-1}.

 \musc\  used HST in combination with X-ray facilities and optical photospheric model atmospheres to obtain 5~\AA\ -- 5.5~$\mu$m spectral energy distributions (SEDs) for seven M dwarf and four K dwarf stars. High-level data products were delivered to MAST as a resource for the exoplanet modeling community as well as stellar astrophysicists to provide a uniform ultraviolet survey for the study of M dwarfs that includes reliable measurements of Ly$\alpha$ and resolved stellar emission lines.  The \musc\ SEDs High Level Science Product archive\footnote{https://archive.stsci.edu/prepds/muscles/} has been used in a number of theoretical studies to model the atmospheric properties of exoplanet atmospheres \citep{spakeetal18-1, wunderlichetal20-1, kawashimaetal19-1}, predict observable atmospheric signatures that can be detected with the JWST \citep{morleyetal17-1}, and investigate the effects of stellar flares on planetary atmospheres and potential life \citep{migueletal15-1, loucaetal23-1} .

The original \musc\ survey focused on stars with known exoplanets. Given the known nearby systems at the time, largely found by optical radial velocity searches, this resulted in a sample that was biased toward the higher mass end of the M dwarf sequence. Only threee \musc\ targets, GJ~1214, GJ~1061 and Proxima Centauri, had masses below $<$0.3~\Msun. With the launch of JWST, the search for ``Earth-like'' planets is moving toward the low mass end of the stellar Main Sequence. Surveys including MEarth, TRAPPIST/SPECULOOS, and TESS are finding Earth-sized planets around stars with masses of $\sim$0.3~\Msun\ or below \citep{bertathompsonetal15-1,dittmannetal17-1,wintersetal19-1,pidhorodetskaetal21-1}. It is targets like these, around low mass stars, that will be the subject of atmospheric characterization of rocky planets in the next decade or more \citep{kemptonetal18-1,fortenbachanddressing20-1, redfieldetal24-1}.

\begin{figure}
    \centering
    \includegraphics[width=\textwidth]{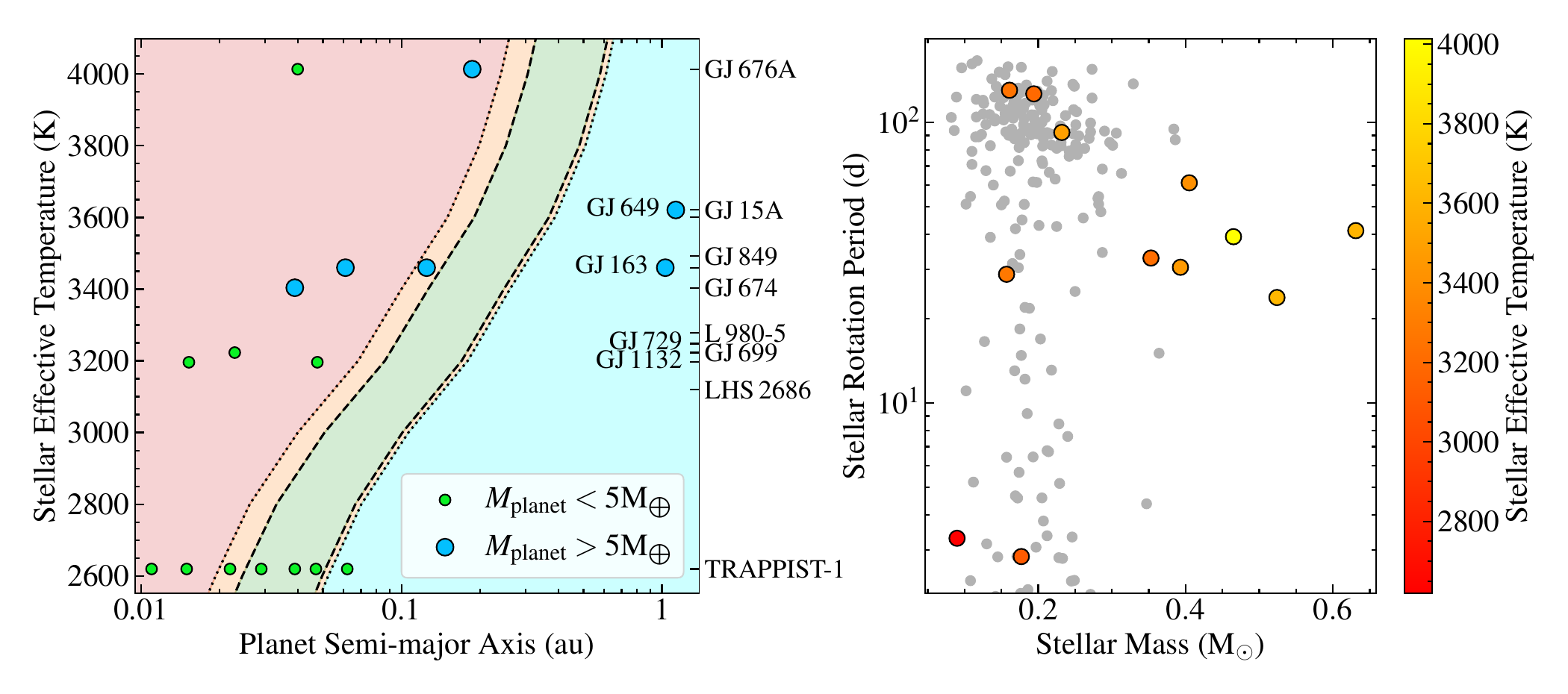}
    \caption{Left panel: Conservative (green) and optimistic (orange) Habitable Zones from \citet{kopparapuetal14-1} as a function of stellar effective temperature for the Mega-MUSCLES sample, along with their planetary systems (Table~\ref{tab:tab_targs}). Note that GJ\,849 b \& c are off the x-axis scale at 2.409\,au and 4.974\, au respectively. Right panel: Stellar masses, temperatures and rotation periods for the Mega-MUSCLES sample. Stars from \citet{newtonetal18-1} are shown in grey for comparison. }
    \label{fig:habzone}
\end{figure}

It was with this in mind that the Mega-MUSCLES Treasury Survey was conceived as the successor to \musc. \megam\ observed thirteen M dwarfs across a range of mass, age, and activity levels (Figure \ref{fig:habzone}), with a particular focus on low mass M stars that are the likely first targets for atmospheric characterization of Earth-like planets. The sample list included observations for six more of the closest low-mass exoplanet host stars; for slow-rotating stars to 0.14~\Msun\ (to serve as proxies for future habitable planet hosts); for stars like GJ~1132 but with faster rotation periods (to track XUV evolution with age); for the 2nd (GJ\,699 = Barnard’s Star) and 7th (GJ\,729) closest star systems; and
for the multiple rocky planet system, TRAPPIST-1. The target list included stars with no known exoplanets that filled out the phase space of M dwarf mass and rotation/activity. When combined with the existing observations, the complete \megam\  library spans a range of stellar masses (0.14--0.8~\Msun), high to low X-ray luminosity fraction (an indicator of activity level), and planetary systems ranging from Jupiters to super-Neptunes to super-Earths.


In this paper, we present an overview of the \megam\ Treasury Survey. Several papers have already been published focusing on individual targets of interest \citep{froningetal19-1,franceetal20-1,wilsonetal21-1}, describing the X-ray observations for both MUSCLES and Mega-MUSCLES in detail \citep{brownetal23-1}, using the observations to probe proxies of stellar coronal and chromospheric behavior \citep{linskyetal20-1}, and developing optical proxies for high energy emission \citep{melbourneetal20-1}. This paper will focus on summarizing the survey observing methods, SED generation, and analysis of the data as a population.

\section{Observations and Data Reduction} \label{sec:obs}

\subsection{Survey Design}
The plan for the original \musc\ survey was to obtain complete spectral coverage of the 1150--3100~\AA\ waveband at moderate spectral resolution of a representative sample of optically inactive M stars with known exoplanet companions. To this end, the team employed the Cosmic Origins Spectrograph \citep[COS,][]{greenetal12-1} and the Space Telescope Imaging Spectrograph \citep[STIS,][]{woodgateetal98-1} onboard HST to obtain ultraviolet and optical spectroscopy using the COS G130M, G160M, and G230L gratings to cover the FUV and faint NUV (1750--2200~\AA) continuum emission and STIS G140M and G230L to observe \lya\ and broadband NUV \citep{franceetal16-1}. They also included STIS G430L mode to obtain observationally inexpensive optical spectra at the same epoch as the ultraviolet observations. Finally, the survey included contemporaneous X-ray observations to provide high energy coverage and additional constraints on estimates of the unobservable EUV emission.  

For \megam, we started with the same observing strategy. However, there were some changes made in response to new observatory policies and restrictions. First, we removed the request for coordinated ultraviolet and X-ray observations given the increasing challenges of scheduling multiple observatories together, particularly HST and Chandra. Second, we relaxed the requirement that all the HST visits be executed within one day of each other to ease scheduling pressure. Since the timescale for M dwarf flares is much shorter than the previous one day gap between the COS and STIS visits, there was no motivation to maintain this requirement. Finally, due to the heightened interest within the community for ultraviolet observations of M dwarfs, HST introduced more stringent policies for clearing M dwarfs under bright object protection (BOP) restrictions. This caused us to migrate the observing modes for several targets from COS G130M/G160M to STIS G140L and, for a few stars, STIS G140L to E140M. For all our targets, we preserved the 5 orbit FUV monitoring observations to establish a baseline on time variability and search for flares.
Tables~\ref{tab:tab_targs} and \ref{tab:params} list the targets observed and their stellar and exoplanet properties, while Table ~\ref{tab_obs} summarizes the ultraviolet and X-ray observations of each target \footnote{The M\,4.5 star LP 756-18 was also observed as part of the Mega-MUSCLES program. Unfortunately the observations failed due to guide star acquisition failures, with only a single STIS G430L exposure returning a usable spectrum (dataset ODLM30010). We therefore do not include LP 756-18 in the Mega-MUSCLES sample or HLSP, but note it here to avoid confusion with, for instance, earlier publications that may mention it as part of the survey.}.

\subsection{Wavelength region definitions}
Throughout the paper we refer to distinct wavelength regions of the SED, the literature definitions of which can vary. Unless specified otherwise, we follow the definitions used for the MUSCLES program by \citet{franceetal16-1}: X-ray: 5--100\,\AA; Extreme Ultraviolet (EUV): 100--911\,\AA; Far Ultraviolet (FUV): 911--1700\,\AA; Near Ultraviolet (NUV): 1700--3200\,\AA\ and optical/IR: $>$3200\,\AA.




\begin{deluxetable}{lccccccc}
\tablecaption{Mega-MUSCLES Target Summary \label{tab:tab_targs}}

\tabletypesize{\small}
\tablecolumns{8}
\tablehead{\colhead{Star} & \colhead{Distance} & \colhead{Spectral Type} &  \colhead{$P_{\mathrm{Rot}}$} & \colhead{Exoplanet Mass} & \colhead{Semi-major Axis} & \colhead{Active?\tablenotemark{a}} & \colhead{Refs} \\
& \colhead{(pc)} &   & \colhead{days} & \colhead{(\mearth)} & \colhead{(au)} }
\startdata
GJ676A & 16.0 & M0 V  & 41.2 & 4.4\tablenotemark{b}, 11.5\tablenotemark{b}, 2127, 2161\tablenotemark{b} & 0.04, 0.187, 1.8152,6.6 &  N & 1 \\
GJ15A & 3.6 & M2 V & 30.5 & 5.35\tablenotemark{b} & 0.072 & N & 2,3, 18 \\
GJ649 & 10.4 & M1 V  & 23.8 & 87.4\tablenotemark{b} & 1.13 & N & 4,5,19,20 \\
GJ674 & 4.6 & M3 V  & 32.9 & 11.1\tablenotemark{b} & 0.039 & N & 6,7,21 \\
GJ729 & 3.0 & M3.5 V  & 2.8 & \nodata & \nodata & Y & 2,3,8 \\
GJ163 & 15.1 & M3.5 V  & 61 & 10.6\tablenotemark{b}, 6.8\tablenotemark{b}, 29.0\tablenotemark{b} & 0.061, 0.125, 1.03 & N & 1,9 \\
GJ1132 & 12.6 & M4 V & 126.6 & 1.66,2.64\tablenotemark{b} & 0.0153,0.0476 & N & 10,11,22 \\
L980-5 & 13.3 & M4 V & 92.2 &\nodata & \nodata& N & 11\\
GJ849 & 8.8 & M3.5 V  & 39.2 & 283\tablenotemark{b}, 342.9\tablenotemark{b} & 2.409, 4.974 & N & 1,5,20,23,24  \\
GJ699 & 1.8 & M4 V & 130.4 & 0.37\tablenotemark{b} & 0.02294 & N & 12,13, 27 \\
LHS2686 & 12.2 & M5 V & 28.8 & \nodata &\nodata & Y \\
TRAPPIST-1 & 12.1 & M7.5 V & 3.3 & 1.017, 1.156, 0.297 & 0.011, 0.015, 0.022 & Y & 14,15,25,26 \\
& & & &   0.772, 0.934 & 0.029, 0.039 & \\
& & & &  1.148, 0.331 & 0.047, 0.062 &  \\
\enddata
\tablecomments{Distances taken from Gaia DR2 \citep{gaia18-1,gaia16-1}.  Rotational period references: 1 \citep{suarezmascarenoetal15-1}, 2 \citep{newtonetal16-1}, 3 \citep{allenandherrera98-1}, 4 \citep{diezalonsoetal19-1}, 5 \citep{veyette+muirhead18-1}, 6 \citep{kiraga+stepien07-1}, 7 \citep{montesetal01-1},  8 \citep{ibanezbustosetal20-1}, 9 \citep{bonfilsetal13-1}, 10 \citep{newtonetal18-1},11 \citep{bertathompsonetal15-1}, 12 \citep{toledopadronetal19-1}, 13 \citep{ribasetal18-1}, 14 \citep{vidaetal17-1}, 15 \citep{burgasser+mamajek17-1}. Exoplanet properties references: 16 \citep{angladaescude+tuomi12-1}, 17 \citep{sahlmannetal16-1}, 18 \citep{howardetal14-1}, 19 \citep{johnsonetal10-1}, 20 \citep{rosenthatetal21-1}, 21 \citep{bonfilsetal07-1}, 22 \citep{bonfilsetal18-1}, 23 \citep{butleretal06-1}, 24 \citep{montetetal14-1}, 25 \citep{gillonetal16-1}, 26 \citep{grimetal18-1}, 27 \citep{gonzalezhernandezetal24-1}. }
\tablenotetext{a}{Optical activity designation where ``Y'' (Yes) indicates targets for which H$\alpha$ is in emission \citep{newtonetal17-1} }
\tablenotetext{b}{$M\sin i$ }
\end{deluxetable}

\begin{table}
    \centering
    \caption{Stellar parameters for the Mega-MUSCLES sample reproduced from Table 3 of \citet{pinedaetal21-1}.}
    \begin{tabular}{l c c c c } \hline \hline
     Name & $L_{\mathrm{bol}}$ (10$^{31}$ erg s$^{-1}$)  & Mass ($M_{\odot}$) & Radius ($R_{\odot}$) & $T_{\mathrm{eff}}$ (K) \\ \hline
	TRAPPIST-1 & $0.234 \pm ^{0.009}_{0.008}$ & $0.090 \pm ^{0.003}_{0.002}$ & $0.120 \pm 0.006 $ & $2619 \pm ^{71}_{66}$ \\ 
	LHS 2686 & $1.078 \pm 0.024$ & $0.157 \pm 0.004$ & $0.182 \pm 0.008 $ & $3119 \pm ^{70}_{68}$ \\ 
	GJ 699  & $1.302 \pm ^{0.024}_{0.023}$ & $0.1610 \pm ^{0.0036}_{0.0035} $ & $0.187 \pm 0.001$ & $3223 \pm 17 $ \\ 
	GJ 729 & $1.537 \pm 0.018$ & $0.177 \pm 0.004 $ & $0.200 \pm 0.008 $ & $3248 \pm ^{68}_{66}$ \\ 
	GJ 1132 & $1.668 \pm ^{0.049}_{0.047}$ & $0.194 \pm 0.005$ & $0.215 \pm 0.009$ & $3196 \pm ^{71}_{70}$ \\ 
	L 980-5 & $2.488 \pm ^{0.079}_{0.078}$ & $0.232 \pm 0.006$ & $0.250 \pm 0.010$ & $3278 \pm ^{74}_{70}$ \\ 
	GJ 674  & $6.03 \pm 0.14 $ & $0.353 \pm 0.008$ & $0.361 \pm ^{0.012}_{0.011}$ & $3404 \pm ^{59}_{57}$ \\ 
	GJ 15A  & $8.608 \pm 0.069$ & $0.393 \pm ^{0.009}_{0.008}$ & $0.385 \pm0.002$ & $3601 \pm ^{12}_{11}$ \\ 
	GJ 163 & $8.28 \pm 0.24$ & $0.405 \pm 0.010 $ & $0.409 \pm ^{0.017}_{0.016}$ & $3460 \pm ^{76}_{74}$ \\ 
	GJ 849 & $11.051 \pm ^{0.095}_{0.094}$ & $0.465 \pm 0.011$ & $0.464 \pm 0.018$ & $3492 \pm ^{70}_{68}$ \\ 
	GJ 649  & $16.74 \pm 0.170 $ & $0.524 \pm 0.012 $ & $0.531 \pm 0.012 $ & $3621 \pm ^{41}_{40}$ \\ 
	GJ 676A & $34.04 \pm ^{0.84}_{0.80}$ & $0.631 \pm 0.017$ & $0.617 \pm ^{0.028}_{0.027}$ & $4014 \pm ^{94}_{90}$ \\ \hline
    \end{tabular}
    \label{tab:params}
\end{table}

\subsection{Ultraviolet and Optical} 

We obtained ultraviolet (1150--3200~\AA) spectroscopy using COS and STIS under HST Program ID 15071. In most cases observations were obtained in photon-counting time-tagged mode to allow for time-resolved light curves and spectra. The spectral resolving power varied depending on observing mode: $R=45,800$ for STIS E140M, $R\sim10,000$ for STIS G140M, $R\sim15,000$ for COS G130M/G160M, $R\sim3,000$ for COS G230L, $R\sim2,000$ for COS G140L, $R\sim1,000$ for STIS G140L, and $R\sim500$ for STIS G230L. 

For each target, we obtained longer (5 orbit) observations in the FUV using either COS G130M (1150--1450~\AA), COS G140L (1150--1700~\AA), or STIS G140L (1150--1700~\AA). The default grating setting for COS G130M was the 1291~\AA\ mode; however, for five targets (GJ676A, GJ649, GJ674, GJ699, and LHS2686), we switched to G130M/1222 to move the BOP-violating  \ion{Si}{4} 1393,1402~\AA\ doublet off the detector; the long cutoff for those monitoring observations is 1365~\AA. For the two stars that were too bright for both COS and the STIS first-order gratings (GJ\,15A and GJ\,729), we used the STIS E140M echelle grating. This had the added benefit of covering the \lya\ line at high resolution. The NUV spectra of those stars were obtained with the STIS/CCD G230LB grating instead of G230L. For stars without E140M observations we used STIS G140M to observe \lya\ at moderate spectral resolution. In addition to the ultraviolet observations, one of the STIS visits for each star included two consecutive short exposures (with the time varying by star) using the CCD G430L setting, covering 2900--5700~\AA\ at a resolving power of R$\sim$500.
 
We extracted light curves from all time-tagged observations to search for stellar flares. For STIS, we used the {\sc stistools}\footnote{\url{https://stistools.readthedocs.io/en/latest/index.html}} inttag routine to split the tag files into 20\,s chunks, then extracted a spectrum from each chunk and integrated it to produce a light curve. For COS,  light curves were extracted from the corrtag event files for each exposure by summing the counts around the location of the spectral trace in the 2-D event list and subtracting background regions of the same size offset in the cross-dispersion direction. 

The spectra were downloaded from the MAST archive after they had been processed using the default {\sc calCOS} and {\sc CalSTIS} data reduction pipelines. For COS observations covering more than one orbit, we shifted each spectrum to the same wavelength scale based on cross-correlation of strong emission lines before coadding, providing a small S/N ratio advantage over the standard x1dsum spectra.    
In four cases (GJ\,1132, GJ\,163, LHS-2686 and TRAPPIST-1) the automated {\sc CalSTIS} pipleline failed to identify the correct spectral trace in some or all of the G140M observations. We re-extracted these spectra using the {\sc stistools} routines, visually identifying the a2center parameter (the y-position of the spectral trace on the detector from the flt files. Once vetted and, if necessary, re-extracted, STIS spectra for each grating were combined using a variance-weighted average.

 All of the HST data from program 15071 can be found in MAST: \dataset[10.17909/jq0n-jh43]{http://dx.doi.org/10.17909/jq0n-jh43}. 

\subsection{X-ray}
\label{sec:xray1}
A full description of the x-ray observations used by MUSCLES and Mega-MUSCLES is presented in \citet{brownetal23-1}, so we briefly summarise here. Six stars were observed with the  ACIS-S3 instrument onboard the Chandra X-ray Observatory (Chandra), five as part of the Mega-MUSCLES program (GJ\,15A, GJ\,163, GJ\,849, GJ\,699, LHS\,2686, program id 19200772 PI Froning), and one retrieved from archival data (L980-5, program 81200661, PI Wright). A further five stars were observed with XMM-Newton (XMM) using the EPIC detectors, with four dedicated observations (GJ\,649, GJ\,674, GJ\,729, TRAPPIST-1, program 081021, PI Froning) and another archival dataset (GJ\,1132, program 080493, PI King). Finally, we retrieved archival observations of GJ\,676A obtained with the XRT on the Neil Gehrels Swift Observatory (Swift). X-ray exposure times for each target are given in Table \ref{tab_obs}.     


\begin{figure}
    \centering
    \includegraphics[width=0.9\textwidth]{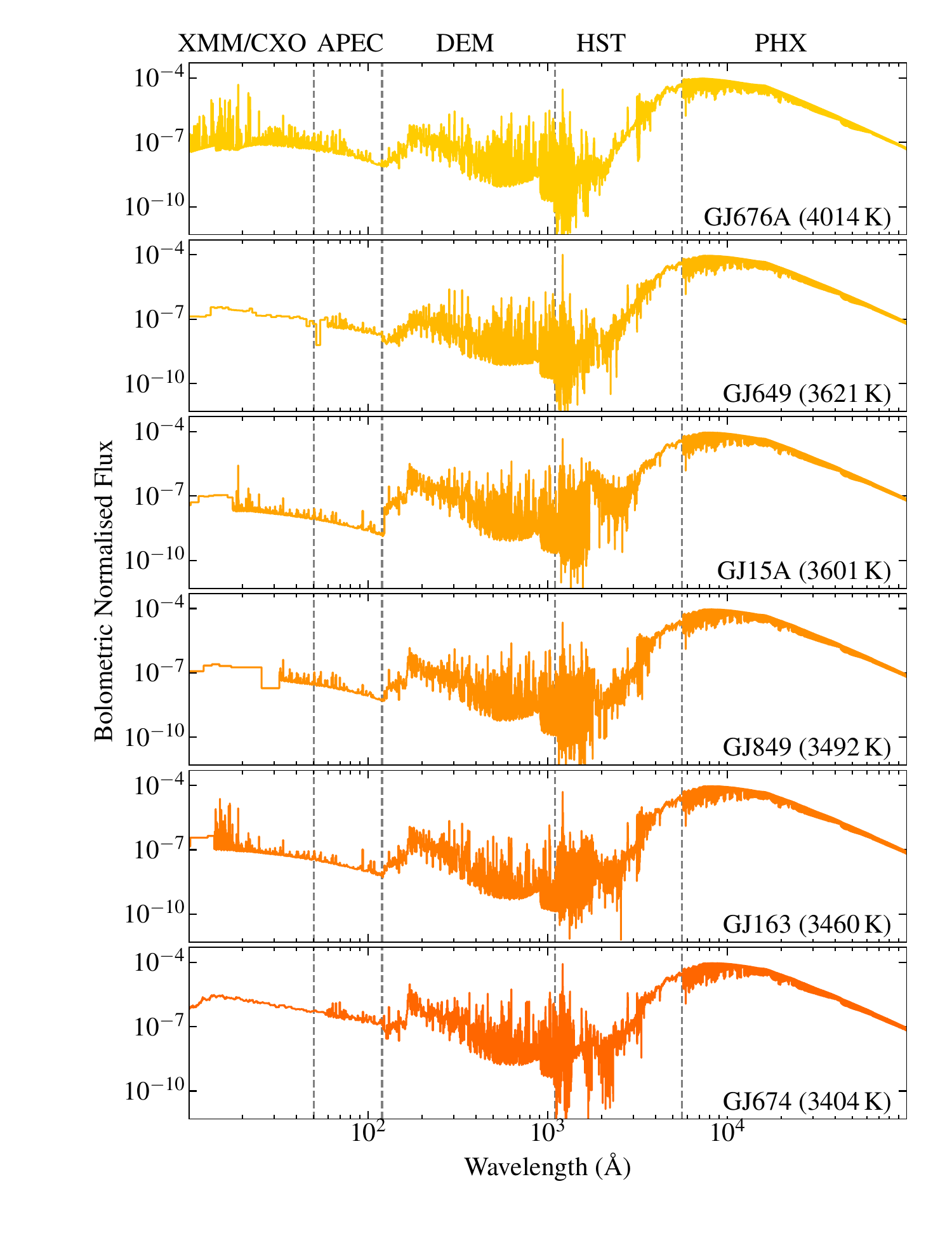}
    \caption{The first six Mega-MUSCLES SEDs in descending order of \Teff. The lines delineating the various data sources are approximate as the exact breaks vary depending on, for example, the S/N of the observed spectra.  }
    \label{fig:seds1-6}
\end{figure}

\begin{figure}
    \centering
    \includegraphics[width=0.9\textwidth]{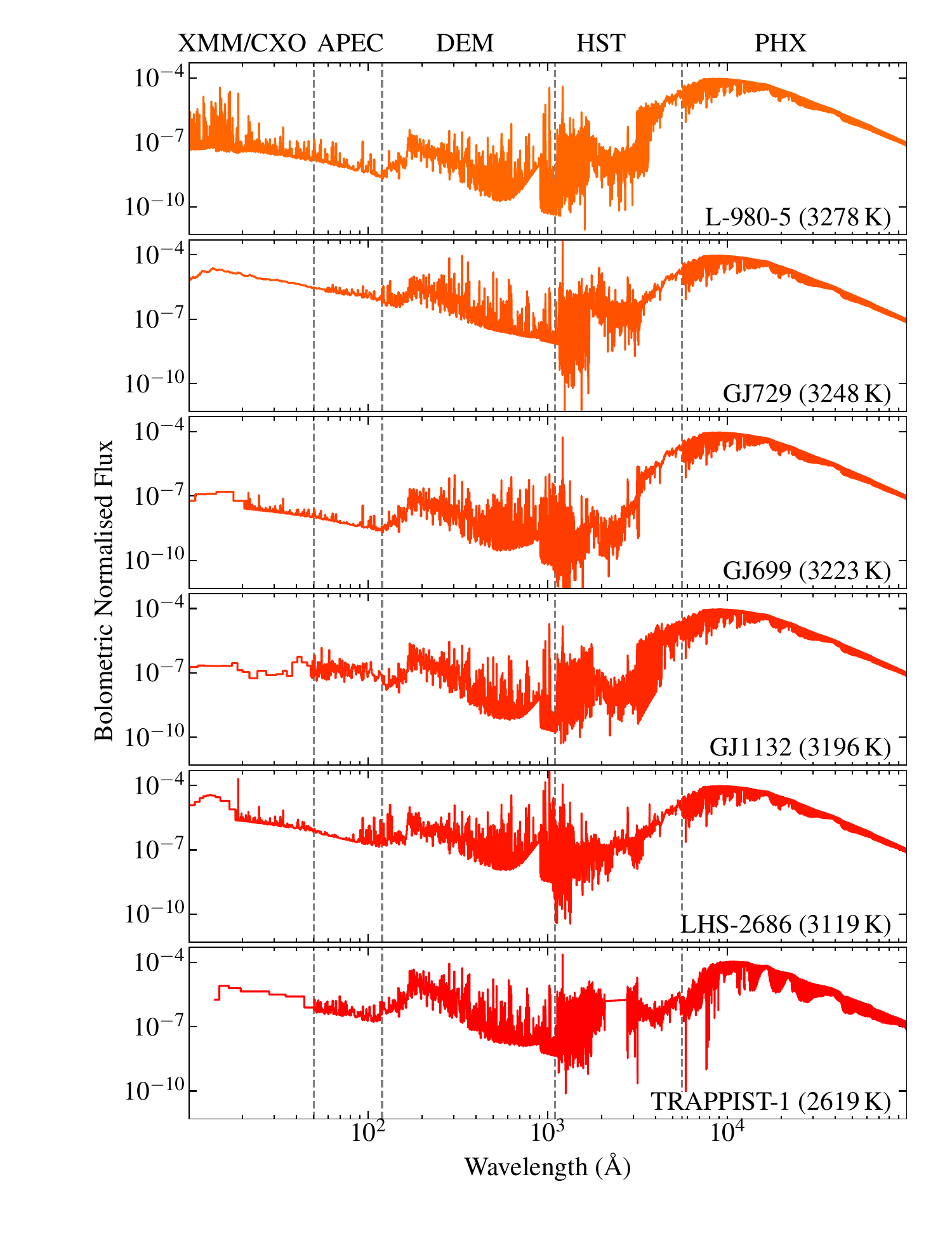}
    \caption{The remainder of the SEDs in descending order of \Teff.}
    \label{fig:seds7-12}
\end{figure}

\section{SED construction} \label{sed}
The construction of the SEDs builds on the techniques used for the MUSCLES sample by \citet{loydetal16-1}, but with multiple changes and improvements. Most notably, we now include estimated uncertainties for the portions of the SEDs filled with models. The final SEDs for all twelve stars are shown in Figures \ref{fig:seds1-6} and \ref{fig:seds7-12}.

\subsection{X-ray}
\label{sec:xray2}
X-ray data were fit using the {\sc XSPEC} package \citep[Versions 12.5.1-12.12.0][]{arnaud96-1}. Each spectrum was fit with a combination of an Astrophysical Plasma Emission Code \citep[APEC,][]{smithetal01-1, fosteretal12-1} model with 1--3 temperatures depending on the quality of the spectrum, modified by absorption from a fixed interstellar hydrogen column density, typically $N_{\mathrm{H}} = 10^{18-19}$\,cm$^{-2}$ (small enough that the effects on the analysis are minimal). The models were then used to fill in the gaps between the end of the X-ray spectra (varied by star) and the lowest wavelengths of the EUV models ($\approx 120$\AA). The XMM spectra were adjusted for photon losses at the low energy boundary as described in \citet{wilsonetal21-1}. In two cases spectra were unavailable: GJ\,676A where only a low count rate from Swift was recovered, and L\,980-5, which was not detected by Chandra. We therefore use APEC models for the entire X-ray region for those stars, fit to the Swift flux measurement and the Chandra upper limits respectively. See \citet{brownetal23-1} for more details.

Uncertainties for the APEC models were generated via a Monte-Carlo approach. For each star, we generated 10,000 models with each parameter randomly drawn from a normal distribution defined by their best fit and 1\,$\sigma$ statistical uncertainty values. This produced a normal distribution for each wavelength bin, from which we take the mean and standard deviation as the final flux and error values respectively.  

\subsection{Extreme Ultraviolet}
The EUV region is currently unobservable for the most part, due to a combination of interstellar hydrogen absorption in the range 400--900\,\AA\ and a lack of sufficiently sensitive instruments at other wavelengths. Mega-MUSCLES SEDs therefore use two different model treatments for the EUV flux. 

The initial release of the Mega-MUSCLES SEDs in early 2022 used the  \citet{linskyetal14-1} empirical scaling relations for the EUV region, whereby the integrated flux of the \lya\ line ($F_{\mathrm{Ly\alpha}}$, Section \ref{sec:lya}) is converted into an EUV flux in 9 wavelength bands, up-sampled to 1\,\AA\ bins. This is the same treatment used in MUSCLES. The only inputs for the EUV scaling relationships are the \lya\ flux and spectral type, so to estimate the uncertainty we generated models with inputs of $F_{\mathrm{Ly\alpha}} + \sigma_{\mathrm{Ly\alpha}}$ and $F_{\mathrm{Ly\alpha}} - \sigma_{\mathrm{Ly\alpha}}$, and used the mean difference of these models from the model generated with $F_{\mathrm{Ly\alpha}}$ as the uncertainty values.

However, these bandpass flux relations are calibrated with a limited sample of stars that are not representative of the broader exoplanet host population. In the updated release accompanying this paper, the EUV regions are instead estimated using Differential Emission Measure models. The ``differential emission measure" technique, or DEM, is an empirically informed but physically constrained method to use observed emission features to constrain the density and temperature structure of the plasma being observed. By assuming that the plasma is collisionally dominated and optically thin, the flux of an emission line can be expressed as the integral of the product of two functions, an ``emissivity" or ``contribution function" that is calculable from atomic data for the line, and an ``emission measure" function which describes the total amount of plasma emitting the line. Many formulations of both functions exist (see \citealt{mariska92-1} for an overview), and the DEM conceptualizes the emission measure as a function of how many collisions along the line of sight excite the line's upper state. The majority of the FUV lines listed in Table \ref{tab:linelist} form over narrow ranges of temperature in the transition region (the \ion{Fe}{12} and \ion{Fe}{21} are coronal exceptions) while the observed X-ray spectrum is largely formed in the corona, making it possible to fit the DEM across a wide range of temperatures. Each measured flux is a localized-in-temperature constraint on what model DEM best fits the data given the calculated contribution function associated with the flux measurement. With a set of likely model DEMs in hand, we can calculate the contribution functions for the emission being formed co-spatially with the the observed features, but at unobserved wavelengths, and calculate the expected flux given the model DEM and calculated contribution function.

\citet{duvvurietal21-1} and \citet{Duvvuri:2023_V1298_Tau} developed an implementation of this technique that propagates both the observational uncertainties and a parameterization of the systematic uncertainty associated with the technique's ability to reproduce the observations forward to the inferred spectrum. These DEM-generated spectra are used for the updated release accompanying this paper and have the advantage of being tailored more specifically to the star using fluxes formed in both the transition region and corona instead of solely the transition region's Lyman$\alpha$ or the coronal X-ray flux. The provision of a spectrum at finer resolution than 100 \AA\ also benefits science cases like the ionization of species other than hydrogen and solving for the population of the excited state of the \ion{He}{1} 10830 \AA\ triplet used to observe atmospheric escape.

The inputs for the DEM are the X-ray spectra/models described above, and measurements of emission line fluxes in the FUV spectra. Table \ref{tab:linelist} provides a summary of emission line fluxes used here, with the number and species of lines dependent on the coverage and S/N of each spectrum. The lines were fitted with Gaussian profiles convolved with the appropriate Line Spread Function\footnote{\url{https://www.stsci.edu/hst/instrumentation/cos/performance/spectral-resolution}}, with Voigt and/or two-component Gaussian profiles used on rare occasions for strong resonance lines. 


\subsection{Far- and Near- Ultraviolet}
FUV and NUV spectra were spliced together, favouring higher resolution data in regions where spectra from different instrument setups overlapped (i.e. favouring COS or STIS E140M spectra). COS spectra are heavily affected by geocoronal airglow from \lya\ and \ion{O}{1}. We conservatively removed \ion{O}{1} regions covering 1300--1310\,\AA\ and 1353--1356\,\AA\ from the COS spectra, replacing them with STIS G140L spectra if available and by a two-degree polynomial fit to the 5\,\AA\ on each side of the gap if not. For \lya\ we removed the region 1207--1222\,\AA\ and replaced it with STIS G140M and the \lya\ reconstruction detailed below. 

\begin{figure}
    \centering
    \includegraphics[width=\textwidth]{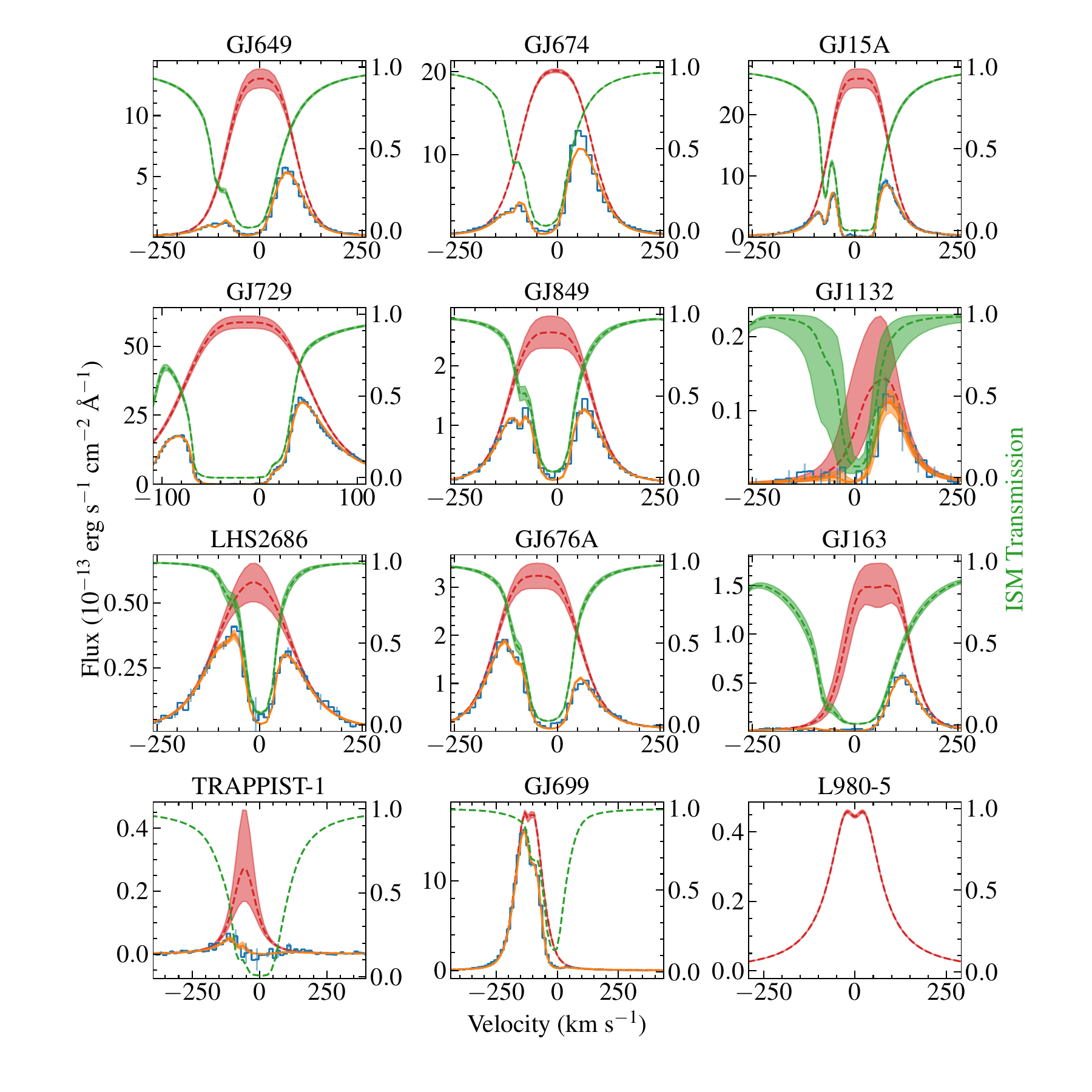}
    \caption{\lya\ profiles for all Mega-MUSCLES stars. Each panel shows the observed G140M or E140M spectrum (blue), the reconstructed intrinsic profile (red), the ISM profile (green, right axis) and the fit to the data, the product of the intrinsic and ISM profiles (orange), all smoothed to their respective instrumental profiles. Shaded regions around each line show the 1\,$\sigma$ uncertainty ranges. The profile for L980-5 is a scaled version of the GJ\,699 profile (see text) so only the intrinsic profile is shown.  }
    \label{fig:lya}
\end{figure}

\begin{table}
    \centering
    \begin{tabular}{lccc}
\hline \hline
Name & $F_{\mathrm{Ly\alpha}}$ (10$^{-14}$ erg s$^{-1}$ cm$^{-2}$ \AA$^{-1}$) & 
$\log$\,N(\ion{H}{1}) (cm$^{-2}$) & Self reversal parameter $p$ \\\hline 
GJ\,1132 & $0.9^{+0.56}_{-0.23}$ & $17.5^{+0.62}_{-0.88}$ & $1.34^{+0.3}_{-0.24}$ \\ 
GJ\,15A & $180^{+8.6}_{-8}$ & $18.1^{+0.021}_{-0.022}$ & $1.02^{+0.036}_{-0.017}$ \\ 
GJ\,163 & $11^{+1.5}_{-1.4}$ & $18.5^{+0.047}_{-0.06}$ & $1.38^{+0.26}_{-0.25}$ \\ 
GJ\,649 & $100^{+4.9}_{-4.6}$ & $18.2^{+0.02}_{-0.021}$ & $1.05^{+0.074}_{-0.036}$ \\ 
GJ\,674 & $170^{+1.4}_{-1.4}$ & $18^{+0.0066}_{-0.0067}$ & $1^{+0.0026}_{-0.0012}$ \\ 
GJ\,676A & $33^{+1.8}_{-1.7}$ & $17.8^{+0.055}_{-0.061}$ & $1.07^{+0.11}_{-0.052}$ \\ 
GJ\,699 & $100^{+2}_{-2}$ & $17.6^{+0.02}_{-0.02}$ & \nodata \\ 
GJ\,729 & $380^{+11}_{-10}$ & $17.7^{+0.029}_{-0.033}$ & $17.72^{+0.03}_{-0.03}$ \\ 
GJ\,849 & $25^{+1.8}_{-1.7}$ & $17.8^{+0.063}_{-0.071}$ & $1.09^{+0.13}_{-0.066}$ \\ 
L\,980-5 & $4.4^{+0.9}_{-0.9}$ & \nodata & \nodata \\ 
LHS-2686 & $5.7^{+0.39}_{-0.38}$ & $17.3^{+0.096}_{-0.13}$ & $0.328^{+0.36}_{-0.23}$ \\ 
TRAPPIST-1 & $1.4^{+0.6}_{-0.36}$ & $18.4^{+0.1}_{-0.1}$ & \nodata \\ \hline
    \end{tabular}
    \caption{Parameters of the \lya\ line reconstructions for the Mega-MUSCLES sample. The values  for GJ\,699 and TRAPPIST-1 were taken from \citet{franceetal20-1} and \citet{wilsonetal21-1} respectively, who used a different formulation for self-reversal which did not provide a $p$ value. L\,980-5 uses a scaled GJ\,699 profile so does not have a measured N(\ion{H}{1}).}
    \label{tab:lyatab}
\end{table}

\pagebreak
\subsubsection[Lyman alpha]{Lyman $\alpha$}
\label{sec:lya}
\ion{H}{1} 1215.67 \AA\ \lya\ is by far the brightest emission line in M dwarf ultraviolet spectra \citep{franceetal13-1}, but \ion{H}{1} gas in the interstellar medium severely attenuates much of the stellar emission. To reconstruct the intrinsic stellar \lya\ flux, we simultaneously fit a model of the interstellar absorption and a model of the intrinsic stellar emission, convolved with the instrument line spread function, to the STIS G140M or E140M spectra. The specifics of the model and fitting process with \texttt{emcee} \citep{foreman-mackeyetal13-1} are described in detail in \cite{youngbloodetal21-1}. In brief summary, we assume Voigt profiles for both the interstellar absorption, stellar emission and self-reversal \citep{youngbloodetal22-1}, fitting a single ISM cloud. Such an assumption is not ideal given the complexity of the local ISM \citep{redfield+linsky08-1}, but is appropriate given that multiple clouds are likely unresolvable with STIS G140M. We relate the \ion{H}{1} and \ion{D}{1} absorbers under standard assumptions that they share the same kinematics, are under thermal equilibrium, and have a fixed abundance ratio (see \citealt{youngbloodetal16-1,youngbloodetal21-1}). Figure \ref{fig:lya} shows the reconstructed lines for each star, together with the underlying data and models, and Table \ref{tab:lyatab} shows the key fitted parameters.

The \lya\ reconstructions for GJ\,699 and TRAPPIST-1 have already been presented in \citet{franceetal20-1} and \citet{wilsonetal21-1} respectively, but we restate the results here for completeness. Additionally, no signal was detected from the target in the G140M observations of L\,980-5, so estimates of the strength of the \lya\ line were made from five strong emission lines elsewhere in the spectrum (\ion{C}{2}\,1335\,\AA, \ion{Si}{4}\,1400\,\AA, \ion{C}{4}\,1550\,\AA, \ion{He}{2}\,1640\,\AA\ and \ion{Mg}{2}\,2800\,\AA) along with the rotation rate, via the scaling relationships from \citet{youngbloodetal17-1}. By averaging the six estimates we find an estimated integrated \lya\ flux of $(4.4\pm 0.9)\times10^{-14}$\,erg\,s$^{-1}$\,cm$^{-2}$. As the surface gravity and spectral type of L\,980-5 are similar to those of GJ\,699, we used the \lya\ profile for Barnard's Star in the L\,980-5 SED, scaled to have the estimated integrated flux and shifted to the correct radial velocity.


The majority of the Mega-MUSCLES \lya\ reconstructions are based on observations with the STIS G140M grating. \citet{wilsonetal22-1} used this grating to observe the \lya\ line of a binary star at different velocities, allowing the line to be observed at different levels of ISM absorption. Reconstructions to those observations returned \lya\ flux differences of up to factor $\sim2$, which they attributed to the inability to properly resolve the \ion{D}{1} line. Further work is needed to properly assess the systematic errors in reconstructions based on G140M, but at present the potential for large imprecision in the \lya\ flux should be taken into account in any study that uses the Mega-MUSCLES SEDs. The reconstructions for GJ\,15A and GJ\,729 are based on E140M data where the \ion{D}{1}\ line is resolved, so should be unaffected by this systematic error.



\subsection{STIS flux normalisation}
For the MUSCLES sample, \citet{loydetal16-1} scaled the fluxes of STIS spectra based on comparisons with overlapping COS data. They found that the STIS data spectra had systematically lower fluxes than the COS data, attributable to slit losses in STIS. For Mega-MUSCLES, we find that significant flux differences between overlapping regions in COS and STIS spectra (either COS G130M compared with STIS G140L or COS G160M compared with STIS G230L) are a) smaller on average than those found for the MUSCLES sample and b) fluxes are higher in STIS, not COS in most cases. The integrated flux from FUV spectra is dominated by emission lines which could genuinely vary between observations, making them an imperfect flux calibrator. Unlike MUSCLES, Mega-MUSCLES has no stars observed with both COS and STIS in the NUV, where the continuum flux is a major contributor. We therefore chose not to scale the Mega-MUSCLES STIS spectra.    

\begin{figure}
    \centering
    \includegraphics[width=7cm]{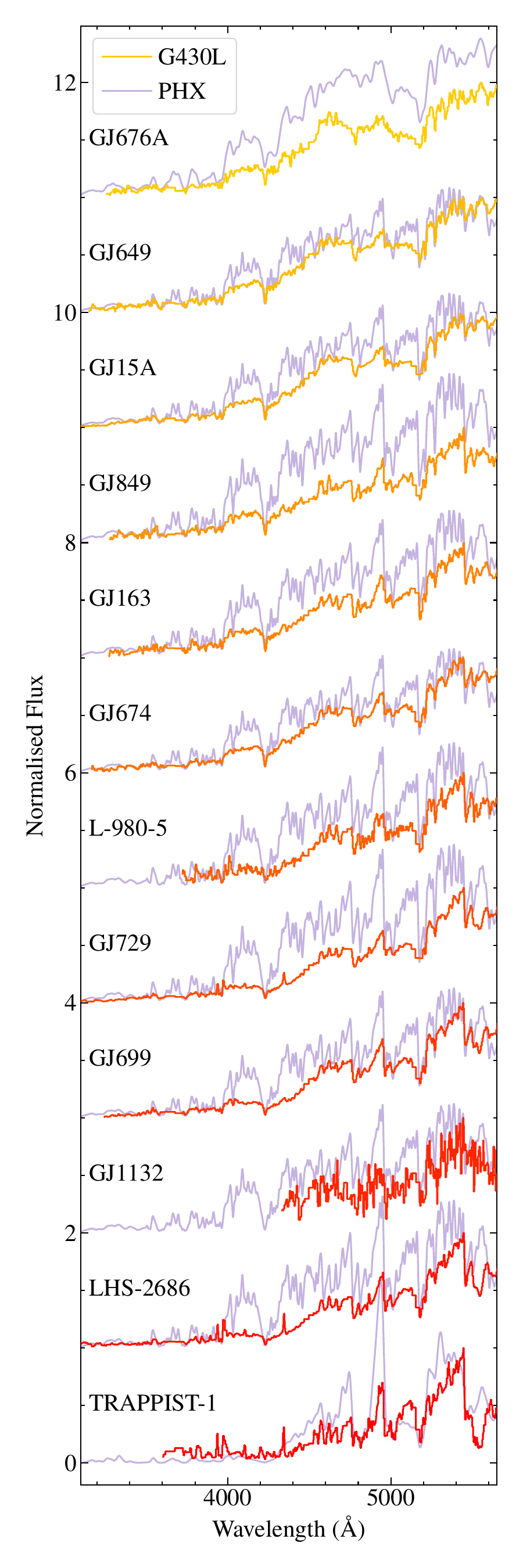}
    \caption{STIS G430L optical spectra for all targets in descending order of \Teff, compared with the PHOENIX (PHX) model used for each star. The mismatch between the peaks of the spectra and models is clearly seen at all stars.}
    \label{fig:g430ls}
\end{figure}

\subsection{Optical}

The blue-optical region of the SED is constructed from two consecutive exposures using the STIS G430L grating, combined in the standard pipeline with a cosmic-ray removal routine. The nominal wavelength coverage of this set up is  2900--5700\,\AA, but in practice the combination of low detector throughput and the decreasing flux from the target at short wavelengths results in an effective non-detection of the blue end of this range at every star. The MUSCLES SEDs used a fixed cutoff at 3850\,\AA\ \citep{loydetal16-1}, but for Mega-MUSCLES we chose to vary the cutoff point for two reasons: In some stars the continuum was clearly detected below this region, and, given the excellent flux calibration (see Section~\ref{subsec:fluxcal}) we wanted to retain as much of the spectrum as possible; in other, fainter stars, only higher wavelength regions were detected.  We therefore removed all points where the mean flux/flux error ratio of the 30 surrounding bins was less than one. The cut-offs for the various stars range from 3067--4334\,\AA. Any gaps between the G230L and G430L spectra were filled with the PHOENIX models described below. The spectra and models are shown in Figure \ref{fig:g430ls}.

\subsubsection{PHOENIX models}
Given the inhomogenety of the available optical to infrared spectroscopy, the SED from 5700\AA\ onwards are filled with a PHOENIX photospheric model spectrum from the Lyon BT-Settl CIFIST 2011\_2015 grid \citep{allard16-1, baraffeetal15-1} retrieved from the SVO theoretical web server service\footnote{\url{http://svo2.cab.inta-csic.es/theory/newov2/index.php?models=bt-settl-cifist}}. The grid models are given as flux at the surface of the star as a function of \Teff\ and \logg. 

Stellar parameters for the Mega-MUSCLES sample were taken from \citet{pinedaetal21-1} and given in Table~\ref{tab:params}. Full details of the measurements are given in that paper, but in short, $K_s$, $J$, and $r$ or $V$ band photometry were combined with the Gaia parallax \citep{gaia18-1} to simultaneously fit the luminosity, mass, radius and \Teff\ of each star via the mass-luminosity relationship of \citet{mannetal19-1}, an empirical mass-radius relationship defined in the paper, and the color-luminosity relationships from \citet{mannetal15-1}. For each star, we used the grid to interpolate a model spectrum for the appropriate parameters, then normalised the spectrum by the squared ratio of the radius and distance. Note that the CIFIST models extend out to wavelengths of 999.5\,$\mu$m, in contrast to the 5.5\,$\mu$m cutoff of the \citet{husseretal13-1} grid used by MUSCLES. 

Given the very large number of data points in the spectrum it was computationally prohibitive to calculate errors in the same way as for the APEC models. By retrieving models using the 1\,$\sigma$ range for each parameter, we found that the uncertainty on \Teff\ had by far the largest impact on the integrated flux ($\approx1\,$percent, compared with 0.01\,percent for the other parameters). Furthermore, uncertainties in radius and distance are propagated though into the \Teff\ uncertainty by \citet{pinedaetal21-1}. We therefore use the mean difference between models for the given \Teff\ and models for \Teff$\pm1\,\sigma$ as a reasonable estimate of the uncertainty.

\begin{figure}
    \centering
    \includegraphics[width=15cm]{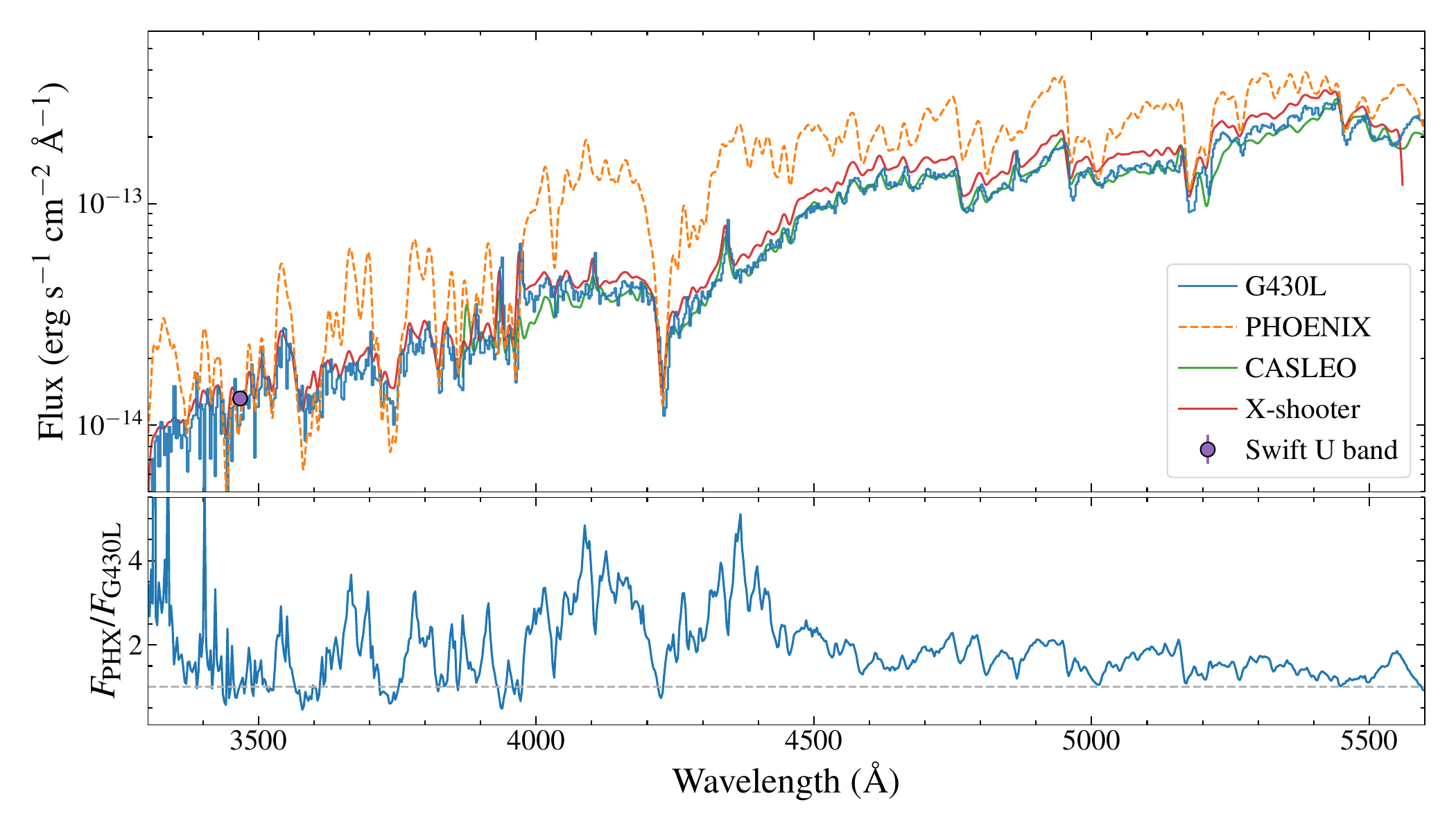}
    \caption{Comparison of blue-optical data of GJ\,729, for which the most data is available from our sources. Top panel: Comparison of STIS G430L, CASLEO and X-shooter spectra, as well as Swift U\,band photometry and the PHOENIX model spectrum. Bottom panel: Ratio between the G430L and PHOENIX spectrum. Similar trends are seen at all other stars.}
    \label{fig:optical}
\end{figure}

\subsubsection{Flux calibration} \label{subsec:fluxcal}
The G430L spectra obtained for MUSCLES were found to systematically under predict the flux from their targets in comparison with broadband photometry, likely due to slit losses due to imperfect target centering in the slit. To attempt to prevent this issue, the Mega-MUSCLES G430L spectra were obtained with a wider slit (0.2 arc\,seconds instead of 0.1), reducing the slit losses. We tested the sucess of this setup by assembling archival flux calibrated spectra from CASLEO and/or X-shooter (Table \ref{tab:archopt}) and Swift U-band photometry. We find that the G430L fluxes obtained from observations with the wider slit are in good agreement with the archival data where available, as shown for GJ\,729 in the top panel of Figure \ref{fig:optical}. We therefore retain the standard pipeline flux calibration for the G430L spectra with no additional scaling.

Figures \ref{fig:g430ls} and \ref{fig:optical} also compare the observed spectra with the PHOENIX models. For all stars, we find that the model matches the depths of strong absorption lines, but overpredicts the broad band continuum by a factor of 2--4. This issue has been noted in several works \citep[see for e.g.][]{fontenlaetal15-1, peraltaetal23-1}, with a general consensus that it is likely due to an incomplete treatment of broad band opacities in the models at blue wavelengths. No applicable solution is currently available, but it should not impact the usability of our SEDs as most of the affected region is covered by the G430L spectra. 

\subsection{Stellar Activity}
M\,dwarfs are notably active stars on multiple timescales (for example, flares on seconds to hours, rotation modulation on days to weeks and stellar activity cycles on months to years), which can strongly affect the X-ray and ultraviolet spectra. Flares can be identified in X-ray and ultraviolet spectroscopy by converting photon-counting/time-tagged data into light curves, and this was done for all suitable observations. We found strong flaring at two stars, each of which is discussed in a separate paper: GJ\,674 \citep{froningetal19-1} and Barnard's Star \citep{franceetal20-1}. At least one small flare was detected in the time-tag data for all stars except for GJ\,1132 and TRAPPIST-1 (the faintest targets). For the large GJ\,674 flare the change in integrated flux between spectra with and without the flare included was $\approx 100$\,percent, but for most flares this difference was much smaller, $\approx5-10$\,percent.

With this in mind, combined with the fact that our X-ray and ultraviolet observations were taken at different times, we have chosen \textit{not} to remove the effects of stellar activity from our SEDs. This is the same approach taken by MUSCLES \citep{loydetal16-1}. The fluxes of the different regions of the SEDs can therefore be only definitively said to be the fluxes of the star, in that waveband, at the particular time of that observation, and might not necessary represent the time-averaged flux of the star. For the stars without flare detections the fluxes can be reasonably assumed to represent the pseudo-quiescent (non-flaring) spectrum of the star, although with the caveat that the X-ray and ultraviolet fluxes might be sampled from different points in the stars rotation and/or activity cycle. \citet{kamgaretal24-1} found that FUV fluxes for 10 M\,dwarfs changed by 30--70\,percent over time scales of decades, although their data was in most cases too sparse to reliably match it to stellar rotation periods or cycles. \citep{loydetal23-1} used extensive FUV observations of the M\,2.5V star GJ\,436 to find a $\approx$40\,percent change in flux due to activity cycles with an $\approx8$\,percent change from rotational modulation. To truly characterise the time-averaged ultraviolet spectra of these stars would require extensive repeat observations, which are unfeasible with our current facilities, but our data and these studies suggest that the contributions from flares, activity cycles and rotation periods are all of order 10s of percent.

\subsection{Final Spectral energy distributions and data products}
Figure \ref{fig:seds1-6} and \ref{fig:seds7-12} show the final assembled SEDs for our 12 targets, the combination of the various spectra and models described in the previous subsections. Our final products are publicly available for retrieval from the MUSCLES High-Level Science Product page hosted by MAST: \url{https://archive.stsci.edu/prepds/muscles/}\footnote{As the updated SEDs will not be available on MAST at the time of the arXiv upload of this paper, we have uploaded them here: \url{https://doi.org/10.5281/zenodo.14081035} }. The archive contains separate spectra for each instrument/grating combination, APEC, DEM, \lya\ and PHX models, as well as the combined SEDs. The SEDs are available in four different formats to provide for a range of user requirements:
\begin{itemize}
    \item {\sc var-res}: The basic product, with all spectra and models included at their native resolutions. We recommend defaulting to this format for modelling.
    \item {\sc adapt-var-res}: This version down samples the observed spectra to remove negative flux points (which are a natural result of backgound subtraction from low S/N spectra). The spectrum is first divided 
    into $\approx$ 10\,\AA\ chunks around prominent emission lines to avoid flux from emission lines being smeared out over a large wavelength range. Within each chunk,  the most-negative flux bin and its two adjacent bins are replaced with the mean flux of all three, repeating until no negative bins remain. Whilst the overall flux is conserved, this does have the effect of smearing out smaller emission lines and can introduce a false continuum. We therefore recommend use of these spectra only when considering the integrated fluxes from large ($\gtrsim 100$\,\AA) wavelength ranges.
    \item {\sc const-res} and {\sc adapt-const-res}: As above, but the SEDs are rebinned to a resolution of 1\,\AA\ using the {\sc specutils Flux Conserving Resampler} routine \citep{Carnall:2017_Flux_Conserving_Resampler}. Rebinning is performed on the individual subspectra before combination into the SED to avoid combining fluxes from different sources into one bin. The native resolution of the PHOENIX models is less than 1\,\AA\, at $\lambda > 1.3\times10^{5}$\,\AA, so bins red-ward of that are removed.  
\end{itemize}
Our aim is that these four formats make the SEDs compatible inputs for as large as possible range of applications. If further custom formats are required please contact the corresponding author. 

\begin{figure}
    \centering
    \includegraphics[width=\textwidth]{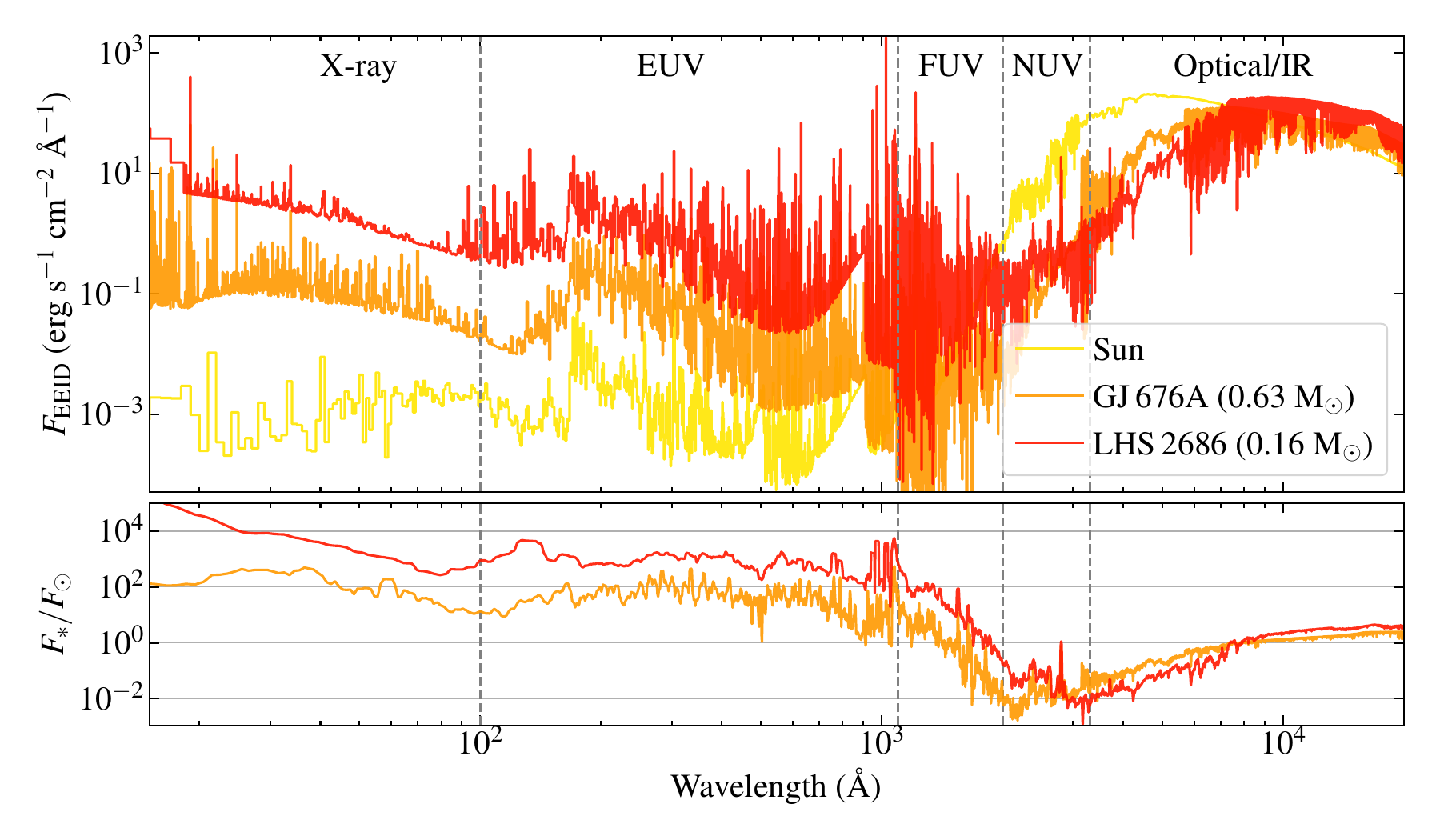}
    \caption{Top panel: SEDs of the hottest (GJ\,676A) and second coolest (LHS\,2686) compared with that of the Sun \citep{woodsetal09-1}. The spectra are scaled to the Earth Equivalent Installation Distance \citep[EEID,][]{mamajek+stapelfeldt23-1}. Bottom panel: The flux ratios of the scaled M dwarf SEDs to the Solar SED.}
    \label{fig:seds_v_sun}
\end{figure}

\section{Discussion}
\subsection{M dwarfs as seen by their planets}
Figure \ref{fig:seds_v_sun} shows the SED of the Sun compared with the hottest and second coolest\footnote{TRAPPIST-1 has already had a paper to itself \citep{wilsonetal21-1}.} stars in our sample. The stellar spectra are scaled to the Earth-Equivalent Instellation Distance \citep[EEID, ][]{mamajek+stapelfeldt23-1}, i.e., the distance at which a planet receives the same total incident flux from the star as the Earth receives from the Sun. The bottom panel shows the ratio of the Solar SED to the M\,dwarfs. It is clearly apparent that the high energy environment faced by M dwarf planets is radically different from that experienced by the Earth \citep{rugheimeretal15-1}. The NUV fluxes are $\sim 100$ times lower but the FUV fluxes are similar, changing the inputs driving photochemical reactions in the upper atmospheres of planets. That is if the planets still have atmospheres: as the XUV fluxes are 100s to 1000s of times higher in M\,dwarf habitable zones than the Earth, planets there will be more susceptible to atmospheric escape \citep{zahnle+catling17-1, vanlooverenetal24-1}. As continued JWST observations define the cosmic shoreline for M dwarf planets, XUV measurements will be vital for predicting and interpreting the presence or absence of atmospheres.            

\begin{figure}
    \centering
    \includegraphics[width=12cm]{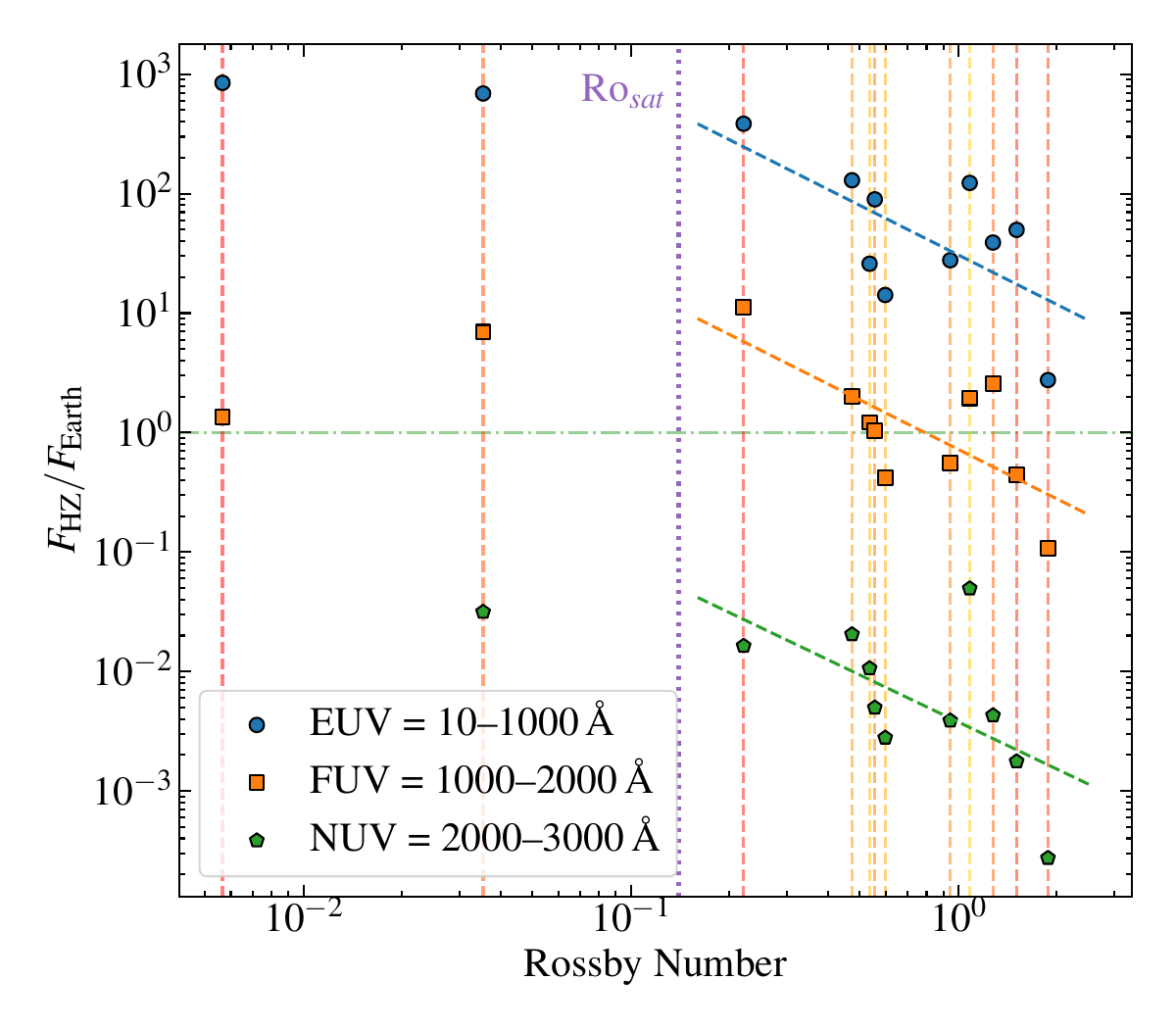}
    \caption{Integrated stellar fluxes for three wavelength ranges experienced in the center of the Habitable Zone for each star compared to the flux received at Earth, as a function of Rossby number. Vertical red/orange lines join the points for each star. Diagonal lines show the power law fits to the data at Rossby numbers higher than the critical Rossby number Ro$_{sat}$ (Ro$_{sat} = 0.14$ from \citet{wrightetal18-1} is shown as a guide, the actual value is approximate and wavelength-dependant.) }
    \label{fig:hz_fluxes}
\end{figure}

Figure \ref{fig:hz_fluxes} quantifies the differences in incident flux in the habitable zone for the entire sample, showing the flux in three wavebands at the EEID for the sample as a function of Rossby number. The Rossby number is the rotation period divided by the convective turnover time, and was calculated via the relationships given in \citet{wrightetal18-1}. The enhancement in EUV flux for planets around M\,dwarfs is readily apparent, and will be even stronger for the closer-in planets that make up the bulk of currently accessible targets for transmission spectroscopy. It is likely that these planets experience significant atmospheric escape over their lifetimes, a conclusion thus far supported by JWST observation of rocky planets \citep[e.g][]{greenetal23-1, moranetal23-1, ziebaetal23-1} which have returned either ambiguous or clear non-detections of thick atmospheres. As atmospheric characterisation of habitable zone planets will take large investments of observing time \citep[e.g.][]{wunderlichetal20-1, linetal21-1}, careful modelling of atmospheric escape at these planets with our SEDs as input should be undertaken to establish if these planets can retain detectable atmospheres. 

As found for multiple other activity tracers, most notably the X-ray luminosity \citep{wrightetal11-1, wrightetal18-1, brownetal23-1}, the ultraviolet fluxes follow a distinct trend with Rossby number: saturation below a certain critical Rossby number Ro$_{sat} \approx 0.1-0.2$ \citep{pinedaetal21-2}, then a power-law decline with higher Rossby number. Only two of our stars are below Ro$_{sat}$ so we cannot confidently infer the flux values for saturation, but the decline above Ro$_{sat}$ is clear. For a power-law decline $F_{\mathrm{HZ}}/F_{\mathrm{Earth}} = a$Ro$^b$ we find:

\begin{equation*}
\mathrm{EUV:}\quad a = 30\pm10,\ b = -1.4\pm0.6\ \quad (R= 0.63) 
\end{equation*}
\begin{equation*}
\mathrm{FUV:} \quad a = 0.7\pm0.2,\ b = -1.4\pm0.5\ \quad (R = 0.72)  
\end{equation*}
\begin{equation*}
\mathrm{NUV:} \quad a = 0.004\pm0.002,\ b = -1.3\pm0.6\ \quad (R = 0.43)
\end{equation*}
With Pearson correlation coefficient $R$. Note is that the slope of the line is identical in all three bands. However there is considerable scatter around the best fit lines with large uncertainties on the fit coefficients, so we can only provide a rough approximation for the  habitable zone high-energy fluxes around any given M\,dwarf: interpreting atmospheric observations ideally  requires either direct observations of the host star as an input, or comparison of the SEDs of several stars with similar spectral types from this and other surveys. \citep{pinedaetal21-2} carried out a similar experiment, fitting individual lines instead of integrated fluxes. They also found that the slope stayed roughly similar with wavelength, but was steeper, ($b\approx-2$ in our notation). With our small sample size and large uncertainties on the fits, it it unclear whether this is a true discrepancy or simple due to small number statistics.




\begin{figure}
    \centering
    \includegraphics[width=12cm]{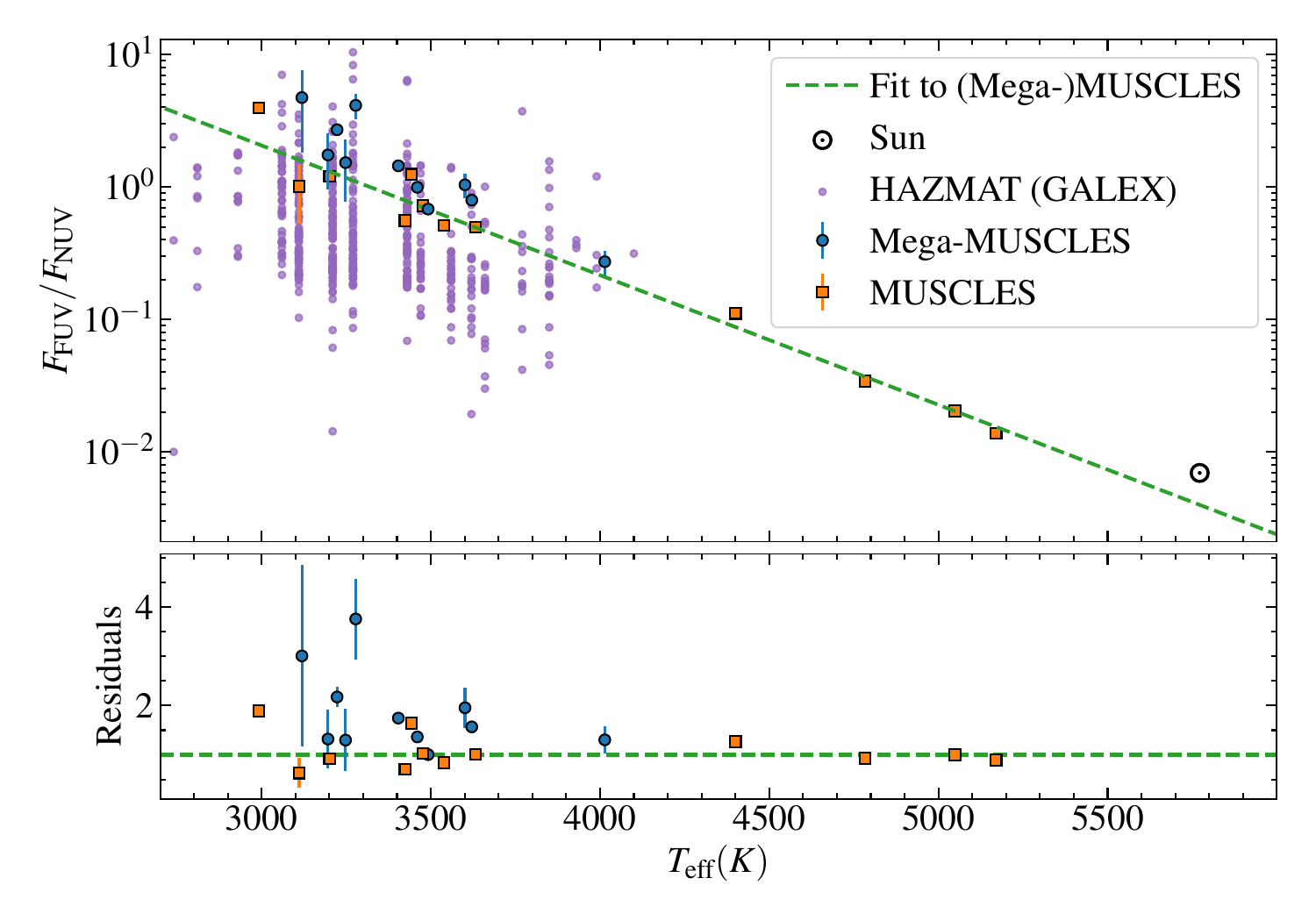}
    \caption{Ratio of the FUV to NUV flux as a function of stellar effective temperature. The bottom panel shows the observed ratios divided by those predicted by the linear fit. GALEX photometry from the HAZMAT project is shown in purple.}
    \label{fig:fnr}
\end{figure}

\subsection{FUV to NUV flux ratio.}
In contrast to the absolute fluxes in the habitable zone, the ratios of the FUV to NUV fluxes of M\,dwarfs follow a distinct trend. The FUV and NUV regions contain peaks in the photoionisation cross-sections of key molecules such as O$_3$ and CH$_4$, and thus the detectable upper atmosphere chemistry of exoplanets will be shaped by the FUV/NUV input it receives from their host stars \citep{migueletal15-1}. Figure \ref{fig:fnr} shows the ratios of the FUV to NUV fluxes for the Mega-MUSCLES and MUSCLES stars \citep{loydetal16-1}, plotted as a function of effective temperature. TRAPPIST-1 was excluded as the NUV signal from the spectra presented here is so weak that special treatment is required to accurately account for the NUV flux \citep{peacocketal19-1, wilsonetal21-1}, and additional NUV observations will be discussed elsewhere (Wilson et al. in prep). Effective temperatures were taken from \citet{pinedaetal21-1} and \citet{loydetal16-1} for the M and K stars respectively. FUV and NUV fluxes were defined as the integral of the ranges 1000--2000\,\AA\ and 2000--3000\,\AA, with small changes in the ranges not affecting the result. As the MUSCLES SEDs do not contain uncertainty estimates for the \lya\ reconstruction or other models, we multiplied the uncertainty on the integrated fluxes by a factor 2 to better compare with the Mega-MUSCLES stars.  

As with previous work \citep{franceetal13-1, behretal23-1}, we find the FUV/NUV ratio falls with increasing effective temperature as the photospheric NUV flux increases. For the temperature range covered by our sample ($\approx$3000--4000\,K) we find that this trend is well fit with a simple linear relationship:
\begin{equation}
\log_{10}(F_{\mathrm{FUV}}/F_{\mathrm{NUV}}) = (-1\pm0.02)\times10^{-3}\,T_{\mathrm{eff}} + (3.3\pm0.1)\quad (R = 0.83)
\end{equation}
The observed ratios are scattered around this line by factors of $\approx0.6-3$ (Figure \ref{fig:fnr}, bottom panel). Whether or not this scatter matters for predictions of exoplanet atmospheric spectra may be an interesting future modelling project \citep{tealetal22-1, cookeetal23-1}. We also note that, in the range $T_{\mathrm{eff}} \approx$3000--4000\,K, FUV/NUV = 1 is a reasonable approximation. The Solar FUV/NUV ratio, calculated using the spectrum from \citet{woodsetal09-1}, falls reasonably close to the line of best fit, although we caution against the use of this relationship for stars at spectral types earlier than K as this trend will turn over at higher temperatures due to increasing encroachment of the photosphere emission into the FUV. Further observations are required to define the FUV/NUV ratio across all spectral types of interest.  

Ideally, spectroscopic observations are required. Figure \ref{fig:fnr} also shows the FUV/NUV ratios calculated from GALEX photometry as part of the HAZMAT survey \citep{schneider+shkolnik18-1}, with spectral type converted into temperature using the table from \citet{pecaut+mamajek13-1}. Whilst they follow the same overall trend, the bulk of the HAZMAT stars fall under the relationship described by the spectroscopic observations. This is likely because the GALEX bands do not cover strong ultraviolet lines, especially \lya, so do not fully account for the stellar FUV and NUV fluxes that an orbiting exoplanet will experience. Whilst photometric observations from GALEX (or the upcoming UVEX mission, which will use similar filters) provide ultraviolet data for a vastly larger number of stars than can be practically obtained with spectroscopy, care must be taken to account for the missing emission features when applying the data to exoplanet atmosphere studies.

\begin{figure}
    \centering
    \includegraphics[width=\textwidth]{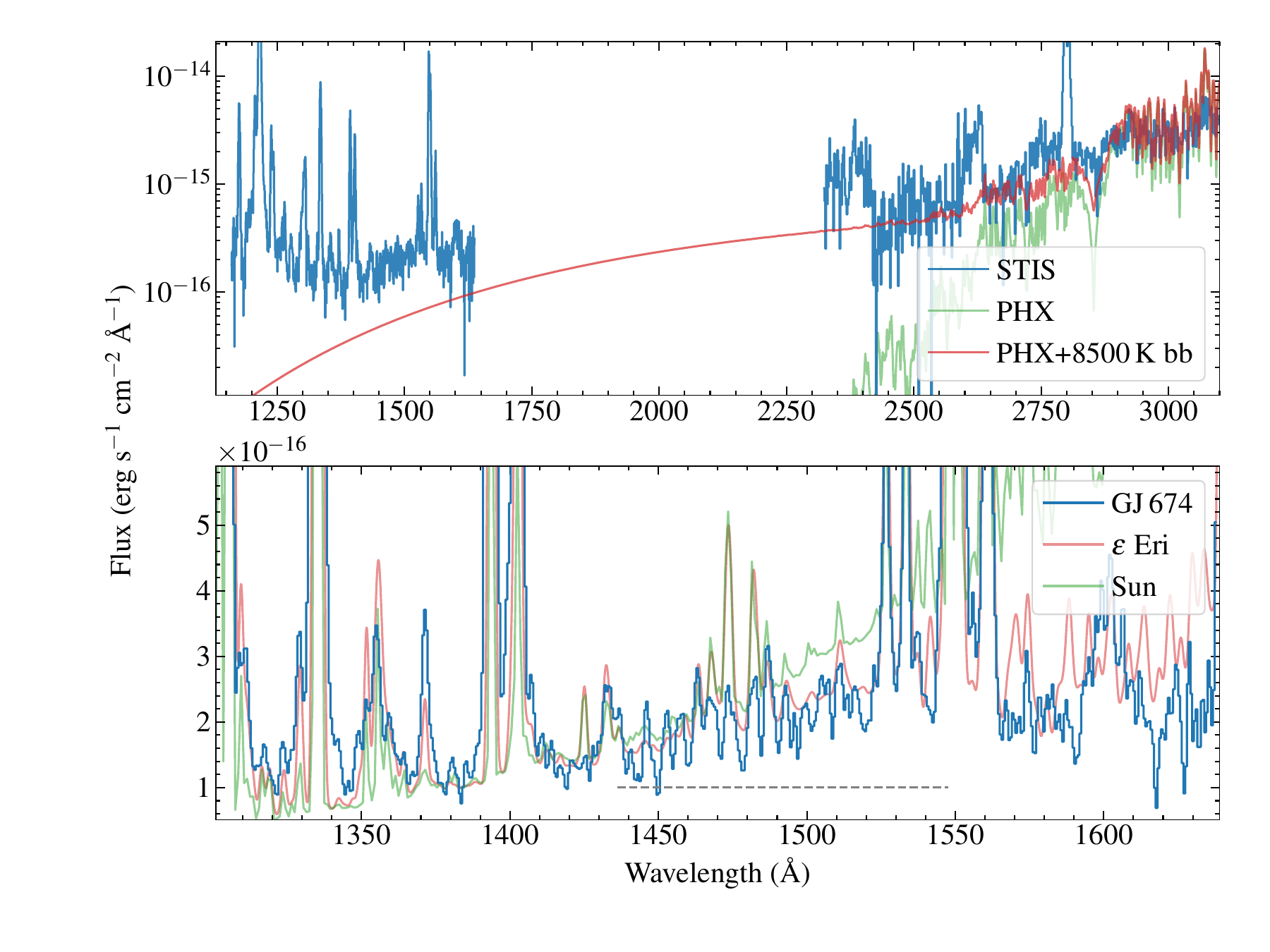}
    \caption{FUV continuum detection at GJ\,674. Top panel: STIS G140L and G230L spectroscopy of GJ\,674. The spectra are compared with the photospheric PHOENIX model used in the SED, and the PHOENIX model combined with a 8500\,K blackbody. Bottom panel: GJ674 compared with $\epsilon$ Eri and the Sun, scaled to have comparable fluxes in the region of interest and convolved to similar resolutions. The dashed line is included to aid the eye to see the potential Si recombination edge.}
    \label{fig:gj_cont}
\end{figure}

\subsection{GJ 674 continuum}
For the majority of the stars in our survey, the regions of the FUV spectra between strong emission lines have low S/N with flux values scattered around zero, i.e., at the noise floor of COS and STIS. The notable exception is GJ\,674, where significant inter-line signal is detected in both COS G130M and STIS G140L spectra (Figure \ref{fig:gj_cont}). The signal in the COS spectrum remains even when the contribution of the large flare described by \citet{froningetal19-1} is removed. The bottom panel of Figure \ref{fig:gj_cont} shows that the inter-line regions are actually a forest of small emission lines, but with underlying structure that is likely due to continuum emission from the stellar atmosphere. As with $\epsilon$\,Eri \citep{loydetal16-1}, there are hints of the \ion{Si}{2} to \ion{Si}{1} recombination edge between 1450\,\AA\ and 1550\,\AA. Unfortunately the range 1700--2250\,\AA\ is not detected in the STIS G230L observation, so we cannot provide a complete description of the stellar continuum across the entire ultraviolet range. However, the top panel of Figure \ref{fig:gj_cont} shows that the common treatment of the chromospheric continuum as a $\approx8500$\,K blackbody \citep{ayres79-1, peacocketal19-1, tealetal22-1} significantly under predicts the stellar flux when extended into the FUV. Deep observations of bright M\,dwarfs coupled with semi-empirical models \citep{tilipmanetal21-1} are required to further understand the FUV continuum.

\section{Mega-MUSCLES survey papers}

As noted in the introduction, there have been several prior publications on individual \megam\ targets, as well as papers using the data for broader survey purposes. Here, we present a brief introduction to the publications, but refer the reader to the papers themselves for more details.

\subsection{TRAPPIST-1}
As a target of critical importance for exopalanetary science, production of the SED of TRAPPIST-1 was prioritised and released as described in \citet{wilsonetal21-1}. The \megam\ TRAPPIST-1 SED remains mostly unchanged from that work, although a few small tweaks have been applied that improve the quality and usability of the high level science products, most notably the addition of uncertainties to the modelled regions of the spectrum.

\subsection{GJ 674} 
During the COS monitoring observations, the hot Neptune planet host GJ~674 exhibited a large stellar flare that persisted over an entire orbit. Although the absolute luminosity of the flare was not high, 
the equivalent duration was $>30,000$~sec, comparable to the Great Flare observed on AD~Leo \citep{hawley+pettersen91-1}. The FUV flare spectrum showed strong continuum emission that was well matched by a blackbody with a color temperature of $T_{br} \approx 40,000\pm10,000$~K, well in excess of typical 9000~K flare emission. \citet{froningetal19-1} presented the flare properties and compared them to parameterizations of radiative hydrodynamic models of chromospheric condensations in stellar flares. They showed that current flare models can only fit the observed continuum with the ad hoc addition of a hot, dense component in the chromospheric emission.
The observation underscored the new information about stellar flare physics that is revealed as more ultraviolet observations of M stars are obtained.

\subsection{Barnard's Star}
Barnard's Star (GJ~699) has a 130~d rotation period and an estimated age of 10 or more Gyr \citep{toledopadronetal19-1}. Despite its classification as an old M dwarf, it exhibited two ultraviolet flares and one X-ray flare during the \megam\ observations, from which \citet{franceetal20-1} estimated a flare duty cycle of 25\%. They used the flare and quiescent spectra to model the effects of the stellar radiation on a hypothetical terrestrial planet in the habiatble zone. The thermal and non-thermal escape models indicated that the quiescent flux was comparable to that seen by the Earth during solar active periods. However, the flare emission can drive significant hydrodynamic mass loss at the estimated duty cycle, suggesting that the recently discovered planet around Barnard's Star \citep{gonzalezhernandezetal24-1}, as well as exoplanets around old M\,dwarfs in general, may still continue to risk significant atmospheric mass loss. 

\subsection{Survey Papers}
The \megam\ observations have been used in conjunction with other ultraviolet survey data in two publications. \citet{linskyetal20-1} used reconstructed \lya\ and X-ray observations from 79 stars to track correlations between stellar chromospheric and coronal emission, respectively. They find trends with stellar type and with age for the G--M sample. The M stars show systematically lower \lya\ emission at a given X-ray flux than the hotter stars. In addition, as stars age they show a smooth increase in normalized (to the bolometric luminosity) \lya\ luminosity to cooler stars at all ages; however, the increase in normalized X-ray coronal emission with decreasing effective temperature is much steeper than that of chromospheric emission in older stars. Thus, in M stars the relative rates of coronal and chromospheric emission vary with age differently than in hotter stars, indicative of changing distribution of heating that favors enhanced coronal emission in cooler stars. 

\citet{melbourneetal20-1} also took advantage of the expanding archival observations of stellar ultraviolet emission to identify optical proxies to high energy emission in M stars. Using archival optical spectra, they examined relations between ultraviolet emission line fluxes and optical H$\alpha$ and \ion{Ca}{2} H\&K fluxes, conducting the analysis as a function of stellar temperature and age. They showed that the relation between ultraviolet line emission normalized to stellar bolometric luminosity and $R^{\prime}_{HK}$ can be used to predict the average ultraviolet emission from a M star in the absence of direct observation to within a factor of 2--4. On the other hand, the normalized ultraviolet luminosities showed no significant correlation with H\,$\alpha$ equivalent width.

\citet{brownetal23-1} presented the details of the X-ray observations used in both MUSCLES and Mega-MUSCLES. The data analysis process used is summarised in Sections \ref{sec:xray1} and \ref{sec:xray2}. They found that the sample falls broadly on to the well-established relationship between X-ray luminosity and rotation \citep{wrightetal18-1}, and that both short and long-term X-ray variability is common, particularly in low mass, fully convective stars.

\section{Summary}
We have presented SEDs for twelve M\,dwarfs, describing our observing strategy, model routines and methodology for each wavelength region. Combined with the original MUSCLES sample, Mega-MUSCLES provides the most comprehensive collection of M\,dwarf SEDs, covering a wide range of spectral types and ages. These data provide a vital input into interpretation of JWST observations of exoplanet atmospheres. Future observations will expand the SED library out to a wider range of spectral types, enabling exoplanet observations across the main sequence with current and future instrumentation.

\acknowledgments
\noindent We thank the anonymous referee for comments that improved the manuscript. 

Based on observations made with the NASA/ESA Hubble Space Telescope, obtained from the Data Archive at the Space Telescope Science Institute, which is operated by the Association of Universities for Research in Astronomy, Inc., under NASA contract NAS 5-26555. These observations are associated with program \# 15071. Support for program \#15071 was provided by NASA through a grant from the Space Telescope Science Institute, which is operated by the Association of Universities for Research in Astronomy, Inc., under NASA contract NAS 5-26555. All of the \textit{HST} data presented in this paper were obtained from the Mikulski Archive for Space Telescopes (MAST). 

This research has made use of data obtained from the Chandra Data Archive, and 
software provided by the Chandra X-ray Center (CXC) in the CIAO application package. 
This research has made use of data from {\it XMM-Newton}, 
an ESA science mission with instruments and contributions directly funded by ESA 
member states and NASA.
This research has made use of data and/or software provided by the High Energy 
Astrophysics Science Archive Research Center (HEASARC), which is a service of 
the Astrophysics Science Division at NASA/GSFC.
Support for X-ray analysis was provided by Chandra grant GO8-19017X.

This work has made use of data from the European Space Agency (ESA) mission
{\it Gaia} (\url{https://www.cosmos.esa.int/gaia}), processed by the {\it Gaia}
Data Processing and Analysis Consortium (DPAC,
\url{https://www.cosmos.esa.int/web/gaia/dpac/consortium}). Funding for the DPAC
has been provided by national institutions, in particular the institutions
participating in the {\it Gaia} Multilateral Agreement.

This research has made use of the NASA Exoplanet Archive, which is operated by the California Institute of Technology, under contract with the National Aeronautics and Space Administration under the Exoplanet Exploration Program.

Y.M. acknowledges funding from the European Research Council (ERC) under the European Union’s Horizon 2020 research and innovation programme (grant agreement no. 101088557, N-GINE). 

%

\vspace{5mm}
\facilities{\textit{HST} (STIS and COS), \textit{XMM-Newton}, \textit{Chandra}}

\defcitealias{astropy18-1}{Astropy Collaboration, 2018}
\software{astropy \citepalias{astropy18-1}, XSPEC \citep{arnaud96-1}, stistools\footnote{\url{https://stistools.readthedocs.io/en/latest/}}, scipy \citep{virtanenetal20-1}, numpy \citep{harrisetal20-1}, matplotlib \citep{hunter07-1}, CHIANTI \citep{Dere:1997_CHIANTI_I_V1, DelZanna:2021_CHIANTI_XVI_V10}}



\appendix
\label{}
\section{Additional Tables}
Here we provide full details of our ultraviolet and X-ray observations for each star (Table \ref{tab_obs}), line lists used to generate the DEMs (Table \ref{tab:linelist}) and a summary of available archival data used to check the flux calibration of the G430L data (Table \ref{tab:archopt}).

\startlongtable
\begin{deluxetable}{ccccccc}
\tablecaption{Observation Summary\label{tab_obs}}
\tabletypesize{\small}
\tablecolumns{7}
\tablehead{\colhead{Star} & \colhead{HST}  & \colhead{Date} & \colhead{T$_{exp}$} & \colhead{X-ray mode} & \colhead{Date} & \colhead{T$_{exp}$}  \\
& \colhead{Mode} & & \colhead{(sec)} & & & \colhead{(ksec)} }
\startdata
GJ676A & COS G130M & 2019-04-05 & 13015 & Swift & 2018-2019 & 6.7 \\
& COS G230L & 2019-04-05 & 333 \\
& STIS G140L & 2019-04-01 & 12595 & \\
& STIS G140M & 2018-06-19 & 5500 \\
& STIS G230L & 2018-06-19 & 3998 \\
& STIS G430L & 2018-06-19 & 2 \\
GJ15A & STIS E140M & 2019-02-13   & 9032 & ACIS & 2019-02-12 & 23.8 \\
& STIS G230LB & 2019-02-12  & 30 \\
& STIS G430L & 2019-02-12 & 3 \\
& STIS E140M & 2019-02-16 & 9032 \\
& STIS G230LB & 2019-02-16  & 30 \\
& STIS G430L & 2019-02-16 &  3 \\
GJ649 & COS G130M & 2018-03-04 & 12638 & EPIC & 2018-03-03 & 18 \\
& COS G230L & 2018-03-04 & 420 \\
& STIS G140L & 2018-03-04 & 6408 \\
& STIS G140M & 2018-03-04 & 3328 \\
& STIS G230L & 2018-03-04 & 100 \\
& STIS G430L & 2018-03-04 & 3 \\
GJ674 & COS G130M & 2018-04-03 & 12947 & EPIC & 2018-04-03 & 30.4 \\
& COS G230L & 2018-04-03 & 341 \\
& STIS G140L & 2018-04-03 & 5486 \\
& STIS G140M & 2018-04-03 & 3243 \\
& STIS G230L & 2018-04-03 & 70 \\
& STIS G430L & 2018-04-03 & 3 \\
GJ729 & STIS E140M & 2018-04-20 & 15200 & EPIC & 2019-03-04 & 23 \\
 & STIS G230LB & 2018-04-19 & 80 \\
 & STIS G430L & 2018-04-19 & 30 \\
GJ163 &  COS G130M & 2018-07-09 & 13133 & ACIS & 2019-30-14 & 28.6 \\
& COS G160M & 2018-07-08 & 10375 & \\
& COS G230L & 2018-07-09 & 2751\\
& STIS G140M & 2019-07-10 & 8251 \\
& STIS G230L & 2019-07-10 & 1556 \\
& STIS G430L & 2019-07-10 & 15 \\
GJ1132 & COS G130M & 2019-04-20 & 11757 & EPIC & 2019-01-10 & 46.6 \\
& COS G160M & 2019-04-17 & 6581  \\
& COS G230L & 2019-04-17 & 5206 \\
& STIS G140M & 2019-04-21 & 5202 \\
& STIS G230L & 2019-04-21 & 4102 \\
& STIS G430L & 2019-04-21 & 4 \\
L980-5 & COS G130M & 2019-03-15 & 12136 & --- & ---\\
& COS G160M & 2019-03-13 & 9542 \\
& COS G230L & 2019-30-13 & 2587  \\
& STIS G140M & 2019-03-17 & 5172 \\
& STIS G230L & 2019-03-17 & 4052 \\
& STIS G430L & 2019-03-17 & 13  \\
GJ849 & COS G130M & 2019-06-09 & 12641 & ACIS & 2019-06-14 & 28 \\
& COS G160M & 2019-06-09 & 6807 \\
& COS G230L & 2019-06-09 & 1011 \\
& STIS G140M & 2019-06-09 & 500  \\
& STIS G230L & 2019-06-09 & 154  \\
& STIS G430L & 2019-06-09 & 3 \\
GJ699 & COS G130M & 2019-03-04 & 12902 & ACIS & 2019-06-17 & 26.7  \\
& COS G230L & 2019-03-04 & 331 \\
& STIS G140L & 2019-03-04 & 7019 \\
& STIS G140M & 2019-03-04 & 5933 \\
& STIS G230L & 2019-03-04 & 205 \\
& STIS G430M & 2019-03-04 & 5 \\
LHS2686 &  COS G130M & 2019-05-31 & 13345 & ACIS & 2019-03-08 & 26.7 \\
& STIS G140L & 2019-06-01 & 5486 \\
& STIS G140M & 2019-06-01 & 5482 \\
& STIS G230L & 2019-06-01 & 1262 \\
& STIS G430L & 2019-06-01 &  283 \\
& STIS G140L & 2019-06-05 & 5486 \\
& STIS G140M & 2019-06-05 & 5482 \\
& STIS G230L & 2019-06-05 & 1262 \\
& STIS G430L & 2019-06-05 &  283 \\
Trappist-1 &  COS G130M & 2018-12-10 & 12404 & EPIC & 2018-12-10 & 24.6 \\
& COS G160M & 2019-06-07 & 1598 \\
& COS G230L & 2017-12-15 & 2731 \\
& STIS G140M & 2018-12-09 & 8131 \\
& STIS G430L & 2018-12-09 &  1795 \\
\hline
\enddata
\tablenotetext{a}{Archival data from other programs.}
\end{deluxetable}

\begin{rotatetable}
\movetableright=0.01mm

\begin{deluxetable*}{lcccccccccc}
\tablecaption{List of FUV emission line flux measurements used to  create the DEMs. Lines marked ``m'' are multiplets where the given wavelength is the approximate midpoint.\label{tab:linelist}}
\tablecolumns{11}
 \tablehead{\colhead{} & \multicolumn{10}{c}{Line Flux (10$^{-16}$ erg s$^{-1}$ cm$^{-2}$)}\\
 \colhead{} & \colhead{Si III} & \colhead{O V} & \colhead{N Vm} & \colhead{Fe XII} & \colhead{S II} & \colhead{Si II} & \colhead{C IIm} & \colhead{Fe XXI} & \colhead{Si IVm} & \colhead{C IVm} \\
 \colhead{Star}& \colhead{1206.499\,\AA} & \colhead{1218.39\,\AA} & \colhead{1240.0\,\AA} & \colhead{1242.0\,\AA} & \colhead{1253.811\,\AA} & \colhead{1309.27\,\AA} & \colhead{1335.0\,\AA} & \colhead{1354.08\,\AA} & \colhead{1400.0\,\AA} & \colhead{1550.0\,\AA}}
\startdata
GJ\,1132 & 0.21$\pm$0.01 & \nodata & 0.25$\pm$0.12 & \nodata & \nodata & \nodata & 0.63$\pm$0.03 & \nodata & 0.07$\pm$0.01 & 1.23$\pm$0.87\\
GJ\,15A & 202.15$\pm$3.41 & 119.8$\pm$1.18 & 294.4$\pm$2.1 & 9.18$\pm$0.45 & \nodata & 33.0$\pm$1.96 & $329\pm3$ & \nodata & 208.10$\pm$6.15 & 1119.30$\pm$8.18\\
GJ\,163 & 6.91$\pm$2.00 & 0.44$\pm$0.03 & 0.63$\pm$0.02 & $<1.6$ & \nodata & \nodata & 1.95$\pm$0.04 & $<$1.50 & $5.6\pm0.28$ & 0.70$\pm$0.04\\
GJ\,649 & \nodata & \nodata & 24.93$\pm$0.17 & \nodata & $1.22\pm0.3$ & \nodata & 44.36$\pm$0.17 & $<$1.73 & $10\pm5.0$ & 65.56$\pm$3.28\\
GJ\,674 & 77.80$\pm$0.37 & \nodata & 121.14$\pm$0.52 & 4.9$\pm$0.16 & $0.87\pm0.06$ & \nodata & 116.90$\pm$0.44 & $<$0.17 & \nodata & \nodata \\
GJ\,676A & \nodata & 5.25$\pm$0.53 & 48.05$\pm$0.23 & \nodata & \nodata & \nodata & 72.10$\pm$7.21 & $<$1.03 &  $14.6\pm1.5$ & 88.44$\pm$8.84\\
GJ\,699 & 14.39$\pm$1.25 & \nodata & 31.49$\pm$0.61 & 0.7$\pm$0.06 & 0.02$\pm$0.16 & 0.35$\pm$0.06 & 50.65$\pm$1.25 & $<$0.24 & \nodata & \nodata\\
GJ\,729 & 28.98$\pm$1.50 & 7.50$\pm$0.69 & \nodata & \nodata & \nodata & \nodata & 34.66$\pm$0.67 & \nodata & $1.94\pm0.8$ & 65.59$\pm$2.08\\
GJ\,849 & $<$0.27 & 5.97$\pm$0.06 & 44.81$\pm$0.21 & \nodata & \nodata & \nodata & 44.73$\pm$0.09 & \nodata & $22.5\pm0.5$ & 33.26$\pm$0.39\\
GJ\,876 & 27.62$\pm$2.67 & 17.38$\pm$1.60 & 47.12$\pm$4.03 & 0.7$\pm$0.07 & $0.37\pm0.06$ & \nodata & 56.41$\pm$5.59 & \nodata & 8.3$\pm$0.8 & 158.88$\pm$12.01\\
L-980-5 & 0.77$\pm$0.05 & $<$1.03 & 0.87$\pm$0.06 & \nodata & \nodata & \nodata & 0.98$\pm$0.02 & \nodata & $1.4\pm0.03$ & 6.30$\pm$4.52\\
LHS-2686 & 7.17$\pm$0.09 & \nodata & 22.26$\pm$0.23 & \nodata & \nodata & \nodata & 16.05$\pm$0.20 & \nodata & \nodata & \\
TRAPPIST-1 & \nodata & \nodata & 0.52$\pm$0.12 & \nodata & \nodata & 0.06$\pm$0.03 & \nodata & \nodata & $0.39\pm0.2$ & 1.53$\pm$0.49\\
\hline
\enddata
\end{deluxetable*}

\end{rotatetable}




\begin{table}
    \centering
    \begin{tabular}{lcc}
    \hline
    \hline
Name & CASLEO & X-shooter \\ \hline
TRAPPIST-1 & N & N \\
L-980-5 & N & N \\
GJ674 & Y & N \\
GJ676A & N & Y \\
GJ649 & N & N \\
GJ699 & Y & N \\
GJ163 & N & Y \\
GJ849 & N & Y \\
GJ1132 & N & N \\
LHS-2686 & N & N \\
GJ729 & Y & Y \\
GJ15A & N & N \\ \hline
    \end{tabular}
    \caption{\label{tab:archopt} Ground based optical spectra used to asses the flux calibration of the G430L spectra.}
    
\end{table}




\bibliographystyle{aasjournal}
\bibliography{aabib,newcites}



\end{document}